\documentclass[a4paper,fleqn,usenatbib]{mnras}

\usepackage{newtxtext,newtxmath}

\usepackage[T1]{fontenc}
\usepackage{ae,aecompl}

\usepackage{epsfig,graphicx}	
\usepackage{amsmath}	
\usepackage{amssymb}	
\usepackage{natbib}
\usepackage{bm}
\usepackage{comment}
\usepackage{dcolumn}
\usepackage{tabu}
\usepackage{datetime}
\settimeformat{hhmmsstime}
\usepackage{multicol}
\usepackage{pdflscape}
\usepackage{subfig}

\title[Jet behaviour in V404 Cygni]{Tracking the variable jets of V404 Cygni during its 2015 outburst
}

\author[A.J. Tetarenko et al.]{A.J. Tetarenko$^{1,2}$\thanks{E-mail: tetarenk@ualberta.ca},
G.R. Sivakoff$^{1}$,
J.C.A. Miller-Jones$^{3}$,
M. Bremer$^{4}$,
\newauthor
K.P. Mooley$^{5,6,7}$,
R.P. Fender$^{5}$,
C. Rumsey$^{8}$,
A. Bahramian$^{3}$,
D. Altamirano$^{9}$,
\newauthor
S. Heinz$^{10}$,
D. Maitra$^{11}$,
S.B. Markoff$^{12}$,
S. Migliari$^{13,14}$,
M.P. Rupen$^{6,15}$,
\newauthor
D.M. Russell$^{16}$,
T.D. Russell$^{12}$,
and C.L. Sarazin$^{17}$
\\
$^{1}$Department of Physics, University of Alberta, CCIS 4-181, Edmonton, AB T6G 2E1, Canada\\
$^2$East Asian Observatory, 660 N. A'ohoku Place, University
Park, Hilo, Hawaii 96720, USA\\
$^{3}$International Centre for Radio Astronomy Research- Curtin University, GPO Box U1987, Perth, WA 6845, Australia\\
$^4$Institute de Radioastronomie Millim\'etrique, 300 Rue de la Piscine, 38406 Saint Martin d'H\`eres, France\\
$^{5}$Department of Physics, University of Oxford, Keble Road, Oxford OX1 3RH, UK\\
$^{6}$National Radio Astronomy Observatory, Socorro, New Mexico 87801, USA\\
$^{7}$Caltech, 1200 E. California Blvd. MC 249-17, Pasadena, CA 91125, USA\\
$^{8}$Astrophysics Group, Cavendish Laboratory, 19 J. J. Thomson Avenue, Cambridge CB3 0HE, UK\\
$^{9}$School of Physics and Astronomy, University of Southampton, Highfield, Southampton SO17 1BJ, UK\\
$^{10}$Astronomy Department, University of Wisconsin-Madison, 475. N. Charter St., Madison, WI 53706, USA\\
$^{11}$Department of Physics and Astronomy, Wheaton College, Norton, MA 02766, USA\\
$^{12}$Anton Pannekoek Institute for Astronomy, University of Amsterdam, P.O. Box 94249, NL-1090 GE Amsterdam, the Netherlands\\
$^{13}$Institute of Cosmos Sciences,
University of Barcelona, Mart\'i i Franqu\`es 1, 08028 Barcelona, Spain\\
$^{14}$XMM-Newton Science Operations Centre, ESAC/ESA, PO Box 78, 28691 Villanueva de la Ca\~nada, Madrid, Spain\\
$^{15}$National Research Council, Herzberg Astronomy and Astrophysics, 717 White Lake Road, PO Box 248, Penticton, BC V2A 6J9, Canada\\
$^{16}$New York University Abu Dhabi, P.O. Box 129188, Abu Dhabi, United Arab Emirates\\
$^{17}$Department of Astronomy, University of Virginia, P.O. Box 400325, Charlottesville, VA 22904, USA\\
}
\date{Accepted XXX. Received YYY; in original form ZZZ}

\pubyear{2018}

\begin{document}
\label{firstpage}
\pagerange{\pageref{firstpage}--\pageref{lastpage}}
\maketitle

\begin{abstract}
We present multi-frequency monitoring observations of the black hole X-ray binary V404 Cygni throughout its June 2015 outburst. Our data set includes radio and mm/sub-mm photometry, taken with the Karl G. Jansky Very Large Array, Arc-Minute MicroKelvin Imager Large Array, Sub-millimeter Array, James Clerk Maxwell Telescope, and the Northern Extended Millimetre Array, combined with publicly available infrared, optical, UV, and X-ray measurements.
With these data, we report detailed diagnostics of the spectral and variability properties of the jet emission observed during different stages of this outburst. These diagnostics show that emission from discrete jet ejecta dominated the jet emission during the brightest stages of the outburst. We find that the ejecta became fainter, slower, less frequent, and less energetic, before the emission transitioned (over 1--2 days) to being dominated by a compact jet, as the outburst decayed toward quiescence. While the broad-band spectrum of this compact jet showed very little evolution throughout the outburst decay (with the optically thick to thin synchrotron jet spectral break residing in the near-infrared/optical bands; $\sim2-5\times10^{14}$ Hz), the emission still remained intermittently variable at mm/sub-mm frequencies. Additionally, we present a comparison between the radio jet emission throughout the 2015 and previous 1989 outbursts, confirming that the radio emission in the 2015 outburst decayed significantly faster than in 1989. Lastly, we detail our sub-mm observations taken during the December 2015 mini-outburst of V404 Cygni, which demonstrate that, similar to the main outburst, the source was likely launching jet ejecta during this short period of renewed activity.
\end{abstract}
\begin{keywords}
black hole physics --- ISM: jets and outflows --- radio continuum: stars --- stars: individual (V404 Cygni, GS 2023+338) --- submillimeter: stars --- X-rays: binaries 
\end{keywords}


\section{Introduction}

Black hole X-ray binaries (BHXBs) contain a stellar mass black hole accreting matter from a companion star, where a portion of the accreted material can be transported back outwards in the form of a relativistic jet. 
These systems are typically transient in nature, evolving from periods of inactivity into a bright out-bursting state lasting days to months.
BHXB jet emission can span many decades in frequency, and during these outbursts, 
the intensity, morphology, spectral, and temporal properties of the jet emission are known to vary with accretion state \citep{fenbelgal04,bel10}.

In quiescence and the hard accretion state, there exists a steady, compact synchrotron-emitting jet, which primarily emits at radio, sub-mm, and optical/infrared (OIR) frequencies \citep{fen01,cor02aa,chat,rus06,tetarenkoa2015,galc05,rus13,plotkin2013,plotkin2015,plotgal15}. This compact jet displays a characteristic flat to slightly-inverted broad-band spectrum ($\alpha\geq0, \hspace{0.1cm}f_{\nu}\propto\nu^{\alpha}$; \citealt{blandford79,falcke95,fen01}) extending up to OIR frequencies  \citep{cor02aa,cas10,chat}, where it breaks to an optically thin spectrum ($\alpha\sim-0.7$; \citealt{rus12}). The location of the spectral break marks the most compact region of the jet, where particles are first accelerated to high energies \citep{mar01,mar05,chat}, and has only been directly observed in a few BHXBs ($\nu_{\rm break}\sim10^{11-14} {\rm \, Hz}$; e.g. \citealt{rus12,rus13}). 

As the system evolves through the rising hard state, where the X-ray luminosity (and, in turn the mass accretion rate) increases, the compact jet spectrum also evolves. In particular, the location of the spectral break has been observed to shift to lower frequencies (toward the radio regime), as the source transitions into a softer accretion state \citep{van13,rus14}. This spectral evolution cannot be driven solely by optical depth effects, which predict an opposite scaling for the spectral break frequency ($\nu_{\rm break}\propto\dot{M}^{2/3}$; \citealt{falcke95}). Alternatively, recent work \citep{kolj15}, which shows a correlation between the location of the spectral break and the photon index of the X-ray spectrum, suggests that the  particle acceleration properties within jets (traced by the flux density and frequency of the spectral break) may instead be connected to the properties of the plasma close to the black hole.

During the transition between the hard and soft accretion state, the jet emission can switch from being dominated by a compact jet to arising from discrete jet ejections (e.g., \citealt{mirbel4,hjr95,kul99,cor02,mj12,brock13,tetarenkoa17}). These ejecta have an optically thin spectrum ($\alpha<0$) above the self-absorption/free-free absorption turnover frequency \citep{mj04}, display highly variable emission, and their expansion/bulk motion can be resolved and tracked with Very Long Baseline Interferometry (VLBI; e.g., \citealt{hjr95,ting95,mj04}). 
Once the source reaches the soft state, jet emission is believed to be quenched altogether (or faint enough to be below the detection thresholds of current instruments; \citealt{fend99a,cor01,russ11b,corr11h,rush16}), with any residual radio emission usually attributed to an interaction between the jet ejecta and the surrounding medium (e.g., \citealt{co04}).  

By tracking spectral, temporal, and morphological changes in the jet emission over an outburst, physical conditions in the jet can be linked to the properties of the accretion flow (probed at X-ray frequencies), potentially revealing which accretion flow properties govern the launching, evolution, and quenching of jets \citep{rus13,rus14,rus15}. 
Therefore, multi-frequency studies of these jets during outburst, which track jet emission properties at different physical scales along the jet axis, are essential in understanding the mechanisms that govern BHXB jet behaviour.
Within the BHXB population, V404 Cygni (aka GS 2023$+$338; hereafter referred to as V404 Cyg) is an optimal candidate for multi-frequency jet studies
due to its proximity (2.39 $\pm$ 0.14 kpc; \citealt{mj9}), low extinction ($E(B-V) = 1.3$; \citealt{cas93}), and high X-ray luminosity levels in outburst ($L_X\sim1\times10^{39}\,{\rm erg\,s}^{-1}$; \citealt{motta17}) and quiescence ($L_X\sim1\times10^{33}\,{\rm erg\,s}^{-1}$; \citealt{corb08}).

V404 Cyg was first discovered
in outburst in 1989 \citep{ma89}, after which it remained in a low-luminosity quiescent state for $\sim26$ years.
During this prolonged quiescent state, V404 Cyg displayed a spectrum that was measured to be flat across the radio band ($\alpha=-0.05\pm0.15$, where $f_\nu\propto\nu^\alpha$ and $\nu=1.4-8.4$ GHz), likely extending up to IR frequencies, where it breaks to an optically thin spectrum ($\alpha<0$; \citealt{hy09}). This radio spectrum is consistent with originating from a partially self-absorbed synchrotron jet \citep{blandford79}. While optical and UV emission in the quiescent spectrum of V404 Cyg are well described by blackbody emission from the known K0IV companion star, \cite{muno6} report a mid-IR ($4.5\mu$m and $8\mu$m) excess above the level expected from the companion star. This mid-IR excess could originate from the accretion disc \citep{muno6,hy09}, the compact jet \citep{gallo7}, or a combination of the two. The average radio flux density of the jet in quiescence is $\sim0.3$ mJy \citep{gallo2003}, although the emission is known to be highly variable (\citealt{hy09}; Plotkin et al. in preparation), reaching up to $\sim 1.5$ mJy \citep{hje00}. 
The quiescent jet of V404 Cyg is unresolved with the global VLBI array, but \cite{mj9} placed an upper limit on the compact jet size scale of $<1.4$ AU at 22 GHz.

During its discovery outburst in 1989, V404 Cyg displayed bright X-ray flaring activity. This highly variable emission was found to not always be intrinsic to the source, but at times be caused by large changes in column density, where the accretion flow became obscured \citep{ter94,oos97,z99}. V404 Cyg displayed a variety of radio behaviour during this outburst \citep{hanhj92}, where the radio spectrum evolved from steep ($\alpha<0$) to inverted ($\alpha>0$) in a matter of days. Further, significant radio flux variability on timescales as short as tens of minutes was observed, and there were hints of coupled radio, optical, and X-ray emission. However, the instruments available during this 1989 outburst did not have the capabilities to perform the simultaneous, multi-frequency, time-resolved observations needed to fully understand this rapidly evolving jet source.

In June 2015, V404 Cyg entered a new outburst \citep{barth15,negoro2015b, kuulkerse15,bern16}, providing a unique opportunity to study the evolving jet with observational coverage that was not possible during the 1989 outburst. During this new outburst, V404 Cyg exhibited bright multi-frequency variability, in the form of large-amplitude flaring events (e.g., \citealt{ferrignoc15,gandhip15,gazeask15,mooleyk15,mottas15a,mottas15b,tetarenkoa15,tetarenkoa15b}), for a $\sim2$ week period, before the flaring activity ceased at all wavelengths, 
and the source began to decay \citep{sivakoffgr15b,sivakoffgr15d,plot16} back towards quiescence. Additionally, in late December 2015 V404 Cyg entered a short mini-outburst period, during which it displayed renewed flaring activity (e.g., \citealt{lip15,trus15,beard15,mal15,tetarenkoa16a,mott16a,mun16b,kaj18}).

In this paper, we present multi-frequency monitoring of V404 Cyg during this 2015 outburst, including radio and mm/sub-mm photometry, combined with publicly available OIR, UV, and X-ray measurements. Radio frequency data were taken with NSF's Karl G. Jansky Very Large Array (VLA) and the Arc-Minute MicroKelvin Imager Large Array (AMI-LA), while the mm/sub-mm frequency data were taken with the Sub-millimeter Array (SMA), the Sub-millimetre Common User Bolometric Array-2 instrument on the James Clerk Maxwell Telescope (JCMT SCUBA-2), and the Institute de Radioastronomie Millim\'etrique's Northern Extended Millimetre Array (IRAM NOEMA).
Our observations span a time period from hours after the {initial X-ray detection of the outburst}, until late in its decay back toward quiescence. While our team's earlier work \citep{tetarenkoa17} probed the jet emission during a portion of the brightest flaring period (on 2015 June 22) of the outburst, this work aims to track the spectral and temporal changes in the jet emission as the system transitioned away from the flaring state and began to decay back into quiescence.
In \S\ref{sec:dr} we describe the data collection and data reduction processes. In \S\ref{sec:res}  we present multi-frequency light curves and broad-band spectra.  In \S\ref{sec:dis},  we use this series of observations to discuss the jet properties in V404 Cyg, as well as draw comparisons to the previous 1989 outburst, and the December 2015 mini-outburst. A summary of our work is presented in \S\ref{sec:sum}.

\section{Observations and Data Analysis}
\label{sec:dr}
\subsection{VLA Radio Frequency Observations}
We observed V404 Cyg with the VLA (project codes 15A-504 and 15A-509) from 2015 July 02 to July 12 (${\rm MJD\,} 57205-57215$) in the L ($1-2\,{\rm GHz}$), C ($4-8\,{\rm GHz}$), Ku ($12-18\,{\rm GHz}$), and K ($18-26\,{\rm GHz}$) bands. The array was in its most extended A-configuration for all observations, where we split the array into 2 or 3 sub-arrays to obtain strictly simultaneous observations across multiple bands. All observations were made with the 8-bit samplers, generating 2 base-bands, each with 8 spectral windows of 64 2 MHz channels, giving a total bandwidth of 1.024 GHz per base-band (see Table~\ref{table:vla_obs} for a summary of the array setup of all the observations). Flagging, calibration, and imaging (with natural weighting chosen to maximize sensitivity) of the data were carried out within the Common Astronomy Software Application package (\textsc{casa} v4.3.1; \citealt{mc07}) using standard procedures outlined in the \textsc{casa} Guides\footnote{\url{https://casaguides.nrao.edu}} for VLA data reduction (i.e., a priori flagging, setting the flux density scale, initial phase calibration, solving for antenna-based delays, bandpass calibration, gain calibration, scaling the amplitude gains, and final target flagging). 
We used J2025$+$3343 as a phase calibrator for all epochs, and 3C48 (0137$+$331) as a flux calibrator in all epochs but July 11 (MJD 57214), where 3C147 (0542$+$498) was used. When imaging the lower-frequency bands ($1-2\,{\rm GHz}$ and $4-8\,{\rm GHz}$), we placed outlier fields on other bright sources within the primary beam to ensure that their side-lobes did not affect our flux density measurement of V404 Cyg. Flux densities of the source were measured by fitting a point source in the image plane (using the \texttt{imfit} task) and, as is standard for VLA L/C/Ku/K band data, systematic errors of 1/1/3/3\% were added \citep{perb17}.  All VLA flux density measurements are reported in Table~\ref{table:vla_flux}.
Given the rapidly changing radio flux density observed in this outburst, we also imaged the source on shorter timescales (less than the full observation period), using our custom \textsc{casa} variability measurement scripts\footnote{These scripts are publicly available on github; \url{https://github.com/Astroua/AstroCompute\_Scripts.}} (see \S 3.1 of \citealt{tetarenkoa17} for a detailed description of the capabilities of these scripts).

\subsection{AMI-LA Radio Frequency Observations}
V404 Cyg was observed with the AMI-LA \citep{zwart2008} radio telescope throughout the 2015 outburst. Observations were carried out with the analogue lag correlator using 6 frequency channels spanning 13.5-18.0 GHz. The raw data were processed (RFI excision and calibration) with a fully-automated pipeline, AMI-REDUCE \citep[e.g.][]{davies2009,perrott2013}.
Daily measurements of 3C48 and 3C286 were used for the absolute flux calibration, which is good to about 10\%. The calibrated and RFI-flagged data were then imported into {\sc casa} for imaging. In this paper, we use a sub-set of the AMI-LA observations taken during this outburst (complete data set will be published in Fender et al., in preparation). Our analysis includes AMI-LA data that were taken simultaneously with our NOEMA mm/sub-mm observations; from 2015 June 26--30 and July 11--12.

\subsection{NOEMA (Sub)-mm Frequency Observations}
We observed V404 Cyg with the NOEMA (project codes S15DE and D15AB) between 2015 June 26 and July 13 (${\rm MJD\,} 57199-57216$), in the 3mm
(tuning frequency of 97.5 GHz) and 2mm (tuning frequency of 140 GHz) bands. These observations were made with the WideX correlator, to yield 1 base-band, with a total bandwidth of
3.6 GHz per polarization (see Table~\ref{table:noema_obs} for a summary of the correlator and array setup of all the observations). We used J2023+336 as a phase calibrator, and MWC349 as a flux calibrator, in all epochs. The bandpass calibrator varied between epochs; 3C273 (57199 at 3mm), 3C454.3 (57200/57203 at 2mm), J1749+096 (57202 at 3mm and 57200/57201 at 2mm) and J2013+370 (57215/57216 at 2mm and 3mm). As {\sc casa} is unable to handle NOEMA data in its original format, flagging and calibration of the data
were first performed in {\sc
gildas}\footnote{\url{http://www.iram.fr/IRAMFR/GILDAS}} using
standard procedures, then the data were exported to {\sc
casa}\footnote{To convert a NOEMA data set for use in {\sc casa}, we followed the procedures outlined at
\url{https://www.iram.fr/IRAMFR/ARC/documents/filler/casa-gildas.pdf}.} for imaging (with natural weighting to maximize sensitivity).  Flux
densities of the source were measured by fitting a point source in the image plane (using the \texttt{imfit} task). All NOEMA flux density
measurements can be seen in Table~\ref{table:submm_flux}. Given the
rapidly changing (sub)-mm flux density observed in this outburst, we
also imaged the source on shorter timescales (less than the full
observation period), using our custom {\sc casa} variability measurement scripts.

\subsection{SMA (Sub)-mm Frequency Observations}
We observed V404 Cyg with the SMA (project code 2015A-S026) between 2015 June 16 and July 02 (${\rm MJD\,} 57189-57215$). All of our observations utilized the ASIC and/or SWARM correlators, tuned to an LO frequency of 224 GHz (see Table~\ref{table:sma_obs} for a summary of the correlator and array setup of all the observations). {We performed all flagging, calibration, and imaging (with natural weighting to maximize sensitivity) of the data within \textsc{casa}, using the same procedures and calibrators outlined in \S 2.2 of \citet{tetarenkoa17}.}
Flux densities of the source were measured by fitting a point source in the image plane (using the \texttt{imfit} task). All SMA flux density measurements are reported in Table~\ref{table:submm_flux}.
Given the rapidly changing (sub)-mm flux density observed in this outburst, we also imaged the source on shorter timescales (less than the full observation period), using our custom \textsc{casa} variability measurement scripts.

\subsection{JCMT SCUBA-2 (Sub)-mm Frequency Observations}
\label{sec:jcmt}
We observed V404 Cyg with the JCMT (project code M15AI54) on 2015 June 17 and July 02 (${\rm MJDs\,\,} 57190$ and $57205$), in the $850\mu$m (350 GHz) and $450\mu$m (666 GHz) bands.  On June 17 the observation consisted of five $\sim30$ min scans on target with the SCUBA-2 detector \citep{chap,holl}, from 11:13:12--14:19:05 UTC (${\rm MJD}\,57190.468-57190.597$). On July 02 the observation consisted of eight $\sim30$ min scans on target with the SCUBA-2 detector from 09:01:23--13:42:29 UTC (${\rm MJD\,}57205.376-57205.571$). 
During the observations on June 17 we were in the Grade 4 weather band with a 225 GHz opacity of 0.1--0.2, while on July 02 we were in the Grade 3 weather band with a 225 GHz opacity of 0.08--0.1. {Data were reduced in the \textsc{starlink} package, using the same procedures and calibrators outlined in \S 2.3 of \citet{tetarenkoa17}.}
JCMT flux densities of the source in both epochs are reported in Table~\ref{table:submm_flux}. We note that we only detect the source at 350 GHz in these epochs; however, $3\sigma$ upper limits in the 666 GHz band are provided in the table.
Given the rapidly changing (sub)-mm flux density observed in this outburst, we also attempted to create maps of the source on shorter timescales (less than the full observation period). To do this, we used a custom procedure we developed to produce a data cube, containing multiple maps of the target source region, at different time intervals throughout our observation (see \S 3.2 of \citealt{tetarenkoa17} for the details of this procedure). In both epochs we were only able to measure the flux density on timescales as short as the 30 min scan timescale, as V404 Cyg was too faint, and the noise was too high to accurately measure the flux density on shorter timescales.

\subsubsection{December 2015/January 2016 Mini-outburst}
\label{sec:jcmt_re}
As V404 Cyg displayed renewed activity in December 2015/January 2016, we also observed V404 Cyg with the JCMT (project code M15BI036) on 2016 January 1 and 2 (${\rm MJDs\,\,} 57388$ and $57389$). Each observation consisted of one $\sim30$ min scan on target with the SCUBA-2 detector (using the daisy configuration), from 19:48--20:20 UTC (${\rm MJD\,} 57388.825$--$57388.847$) on January 1, and 19:59-20:32 UTC (${\rm MJD\,} 57389.833$--$57389.856$) on January 2. CRL2688 was used for absolute flux calibration on January 1, and Mars was used on January 2. During both observations we were in the Grade 1 weather band, with a 225 GHz opacity of 0.04/0.05 on January 1/2. These later epochs were also reduced in the \textsc{starlink} package, following the same procedures as in \S\ref{sec:jcmt}. V404 Cyg transits during the daytime at this time of year, therefore, the JCMT was operating in an specialized extended observing mode at the time of our observations.
As we observed the (sub)-mm flux density change on rapid timescales during the main outburst of V404 Cyg, we opted to also search for variability within the January 1 observation by splitting the scan into two maps. The first half of the observation shows an average flux density of $58\pm19\,{\rm mJy}$ and the second half shows an average flux density of $38\pm10\,{\rm mJy}$. The source was not bright enough, and the noise was too high to accurately measure the flux density on shorter timescales. For the same reasons we were unable to obtain flux density measurements on timescales less than the 30 min scan timescale in the January 2 epoch. 
JCMT flux densities of the source in these later epochs are also reported in Table~\ref{table:submm_flux}.

\subsection{IR/Optical/UV/X-ray Observations}
We have compiled publicly available OIR, UV, and X-ray ({{\em Chandra}; \citealt{plot16}, and {\em Swift/XRT}; \citealt{sivakoffgr15d}}) photometric observations that were quasi-simultaneous with our radio through sub-mm observations (i.e., $<$ 1 day separation from our observations). 
Observational details and flux densities from these data are reported in Table~\ref{table:other_flux}, where data in this table have been de-reddened (when required) using the prescription in \cite{cardelli89}, with an $E(B-V)=1.3\pm0.2$ \citep{cas93}.
Additionally, in our analysis we include time-resolved OIR data from \citealt{kim16} and AAVSO\footnote{Kafka, S., 2018, Observations from the AAVSO International Database, \url{https://www.aavso.org}}, as well as
INTEGRAL X-ray data ({3--10 keV and 60--200 keV bands from the JEM-X and ISGRI instruments, respectively}; \citealt{rod15aa})\footnote{All INTEGRAL X-ray data presented in this paper are taken from the INTEGRAL public data products available at \url{http://www.isdc.unige.ch/ integral/analysis\#QLAsources} (\citealt{kuulkerse15c}, PI: Rodriguez).}, all occurring simultaneously with our radio/sub-mm data sets.

{The {\em Swift/XRT} data were all taken in photon counting (PC) mode, and we analyzed the data using the {\sc heasoft} software package. We first reprocessed the data using \texttt{xrtpipeline}, and then we extracted source and background spectra using standard procedures in \texttt{xselect}. Due to the presence of dust scattering halos around V404 Cyg in many of these observations (e.g., \citealt{beard2016,sebh16}), background spectra were extracted from regions where no halos were detected. 
We fit the 0.5--10 keV X-ray spectra with an absorbed power-law (\texttt{TBABS}*\texttt{PEGPWRLW}); here we used abundances from \citet{wilms00} and cross sections from \citet{vern96}. We tied the power-law photon index ($\Gamma=1.8\pm0.3$; 90\% confidence interval) together, but allowed the hydrogen absorption column to vary between observations ($N_H\sim0.5\mbox{--}3\times10^{22} {\rm \, cm^{-2}}$).
We report the flux densities arising from these fits at 5 keV ($1.21\times10^{9} {\rm \, GHz}$). 
}
 \begin{figure}
\begin{center}
 \includegraphics[width=1\columnwidth]{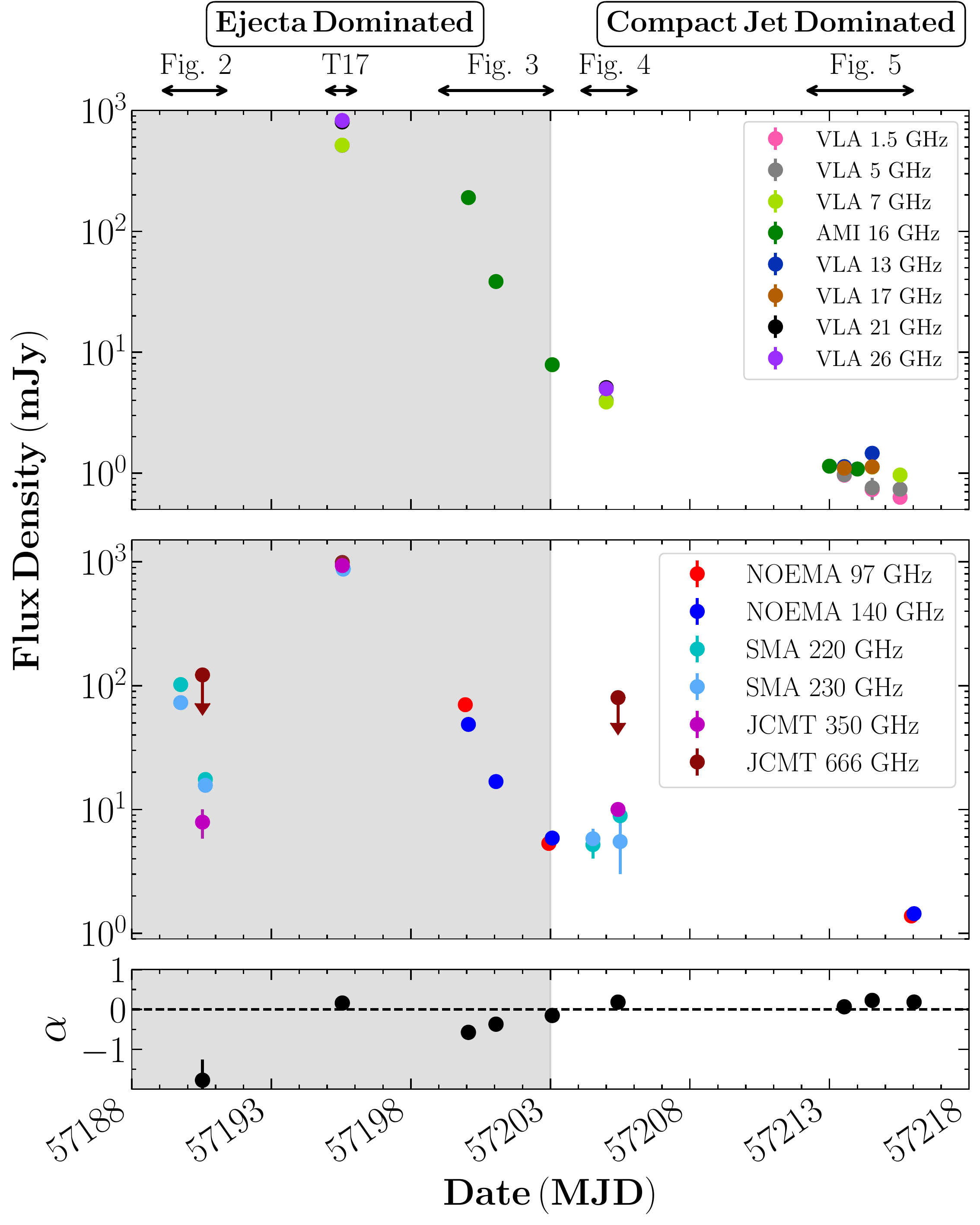}
 \caption{\label{fig:dailylc}   Day-timescale light curves of V404 Cyg during its June 2015 outburst. The {\sl top} panel displays the radio frequency bands, the {\sl middle} panel displays the (sub)-mm frequency bands, and the {\sl bottom} panel displays the radio--sub-mm spectral indices, in epochs where at least two different bands were sampled (using the convention $f_\nu\propto\nu^\alpha$, where $\alpha$ represents the spectral index; dotted line indicates $\alpha=0$). The (un-)shaded regions represent time-periods where the jet emission was likely dominated by jet ejecta or a compact jet (as labelled above the top panel). The figures displaying time-resolved measurements of these data are indicated at the top of the figure (T17 indicates \citealt{tetarenkoa17}). Over our month long monitoring period of V404 Cyg, we find that the jet emission is highly variable, where the radio through sub-mm fluxes vary by $\sim3$ orders of magnitude, and the spectral index varies between steep ($\alpha<0$) and inverted ($\alpha>0$). 
}
\end{center}
\end{figure}

\begin{figure}
\begin{center}
 \includegraphics[width=1\columnwidth]{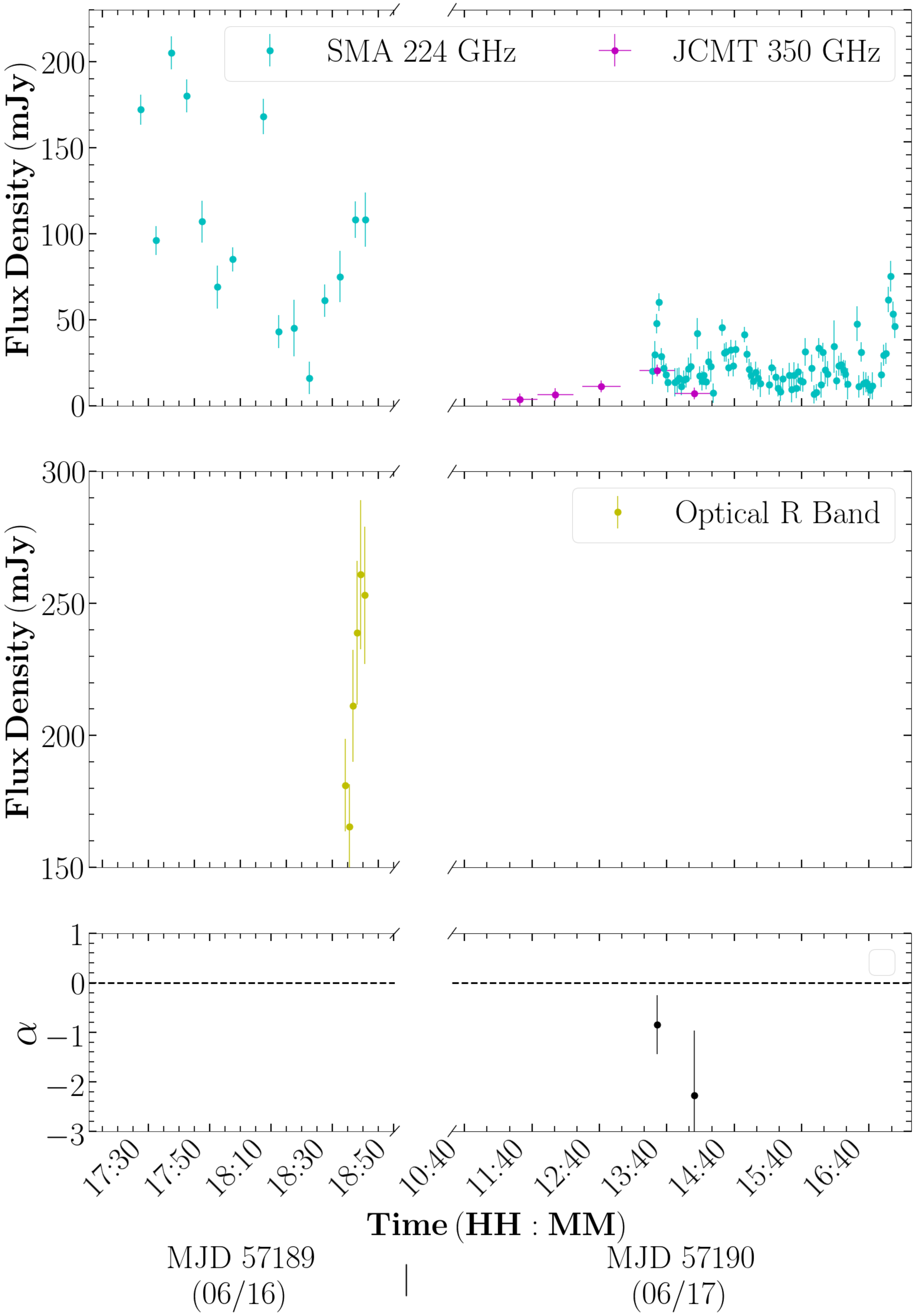}
 \caption{\label{fig:trlc_jun1617}  Time-resolved light curves and spectral indices of V404 Cyg during the first two days of our monitoring of the outburst (June 16 and 17, MJDs 57189 and 57190). The data shown have varying time-bin sizes; 224 GHz (5/2 min June 16/17), 350 GHz (30 min), Optical R band (75 sec; \citealt{kim16}). Here we have combined the two SMA sidebands (cyan data points), in order to gain a higher signal to noise in our time-resolved light curves. The horizontal error bars on the JCMT measurements (magenta data points) represent the time range of the 30 minute SCUBA-2 scans. The dotted line in the bottom panel indicates a spectral index of $\alpha=0$.  In less than 24 hours, between these two epochs, both the sub-mm flux levels and variability amplitude change dramatically.
}
\end{center}
\end{figure}

 \begin{figure}
\begin{center}
 \includegraphics[width=1\columnwidth]{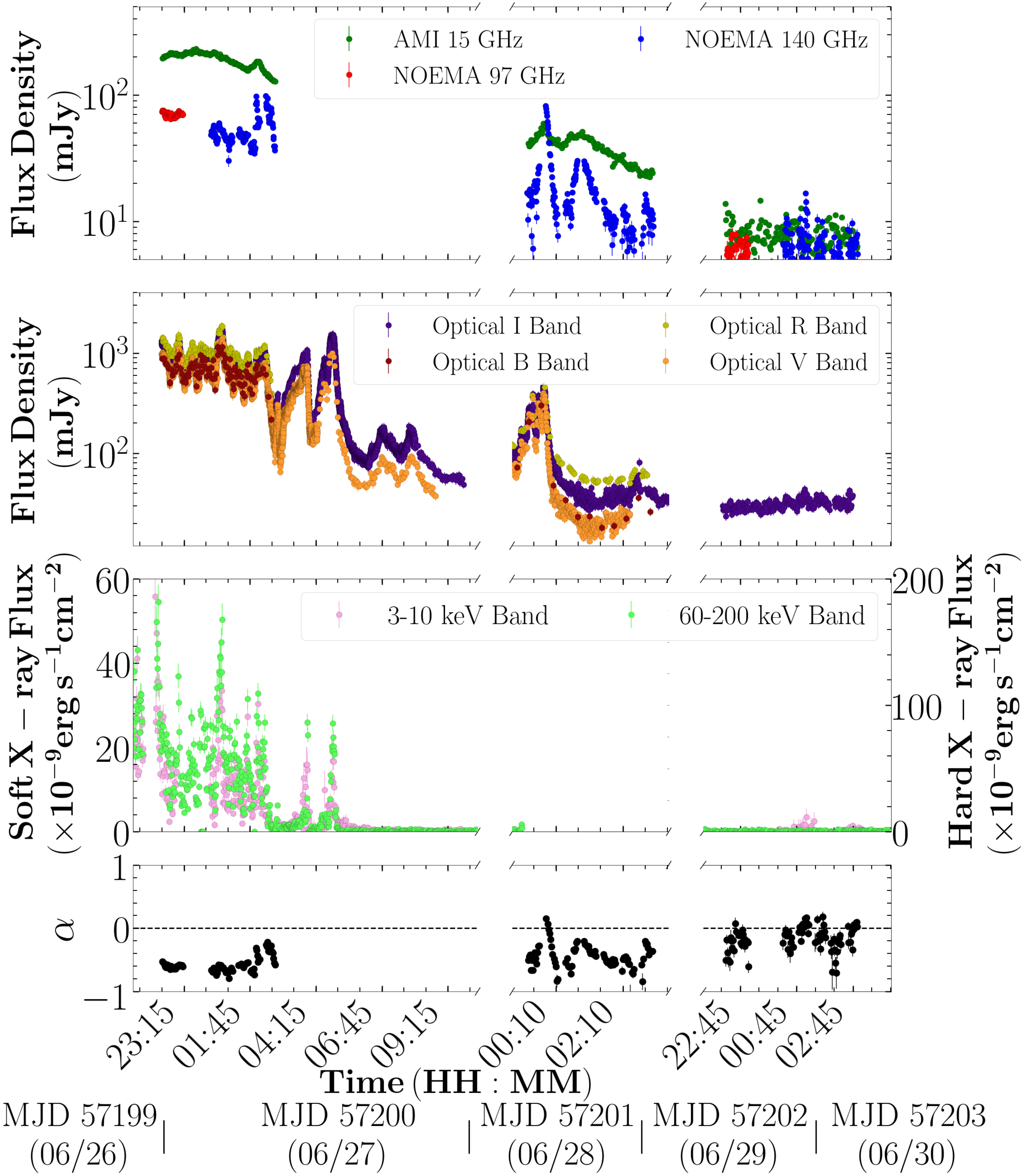}
 \caption{\label{fig:trlc_jun2630} Time-resolved light curves and spectral indices of V404 Cyg between June 26 and 30 (MJD 57199--57203). The data shown have varying time-bin sizes; 15 GHz (100 sec), 97-140 GHz (45 sec), Optical I, R, B,V bands (75 sec; \citealt{kim16}), INTEGRAL 3-10 keV (64 sec), INTEGRAL 60-200 keV (64 sec). The dotted line in the bottom panel indicates a spectral index of $\alpha=0$. We initially detect rapid flaring activity at radio through X-ray frequencies, which drops in amplitude, becomes less frequent, and eventually stops all together, over this four day period.
}
\end{center}
\end{figure}

\section{Results}
\label{sec:res}
\subsection{Light Curves}

Daily timescale light curves of all of our radio through sub-mm observations of V404 Cyg are presented in the top two panels of Figure~\ref{fig:dailylc}\footnote{All the fluxes presented in this figure are measured through imaging the source over the full observation period, except for the AMI 16 GHz data, where a weighted mean of time-resolved measurements (100 sec time-bins) is taken.}. 
Throughout our month long-monitoring of the source, we observe the radio/sub-mm flux to vary by over 3 orders of magnitude, ranging from Jy levels at its brightest to sub-mJy levels at its faintest. In the sub-mm bands, the emission in our first epoch (taken hours after the first detection of the outburst in X-rays) is relatively bright compared to the mm/sub-mm flux densities typically seen in BHXBs (i.e., $\sim 100$ mJy vs $<50$ mJy), but rapidly drops by an order of magnitude within the next 24 hours.
Following our first two detections, the sub-mm flux likely continues to rise, approaching a peak on MJD 57195. The source then begins to decay, where this decay is initially quite rapid (i.e., the flux density drops at least an order of magnitude between MJD 57195 and 57199), before the emission appears to plateau for a few days around MJD 57204, and then proceeds to decay at a much slower rate as the source heads towards quiescence. The radio emission tracks the sub-mm emission closely, and both show a potential secondary peak in the light curves around MJD 57200.

The bottom panel of Figure~\ref{fig:dailylc} displays the radio through sub-mm spectral indices (where a single power-law is fit across radio/sub-mm frequencies), for epochs where at least 2 different bands were sampled. We find that the spectral indices appear to vary between steep ($\alpha<0$) and inverted ($\alpha\geq0$) during the MJD 57189--57204 period, but then remain flat to inverted for the rest of our monitoring period.

\begin{figure}
\begin{center}
 \includegraphics[width=1\columnwidth]{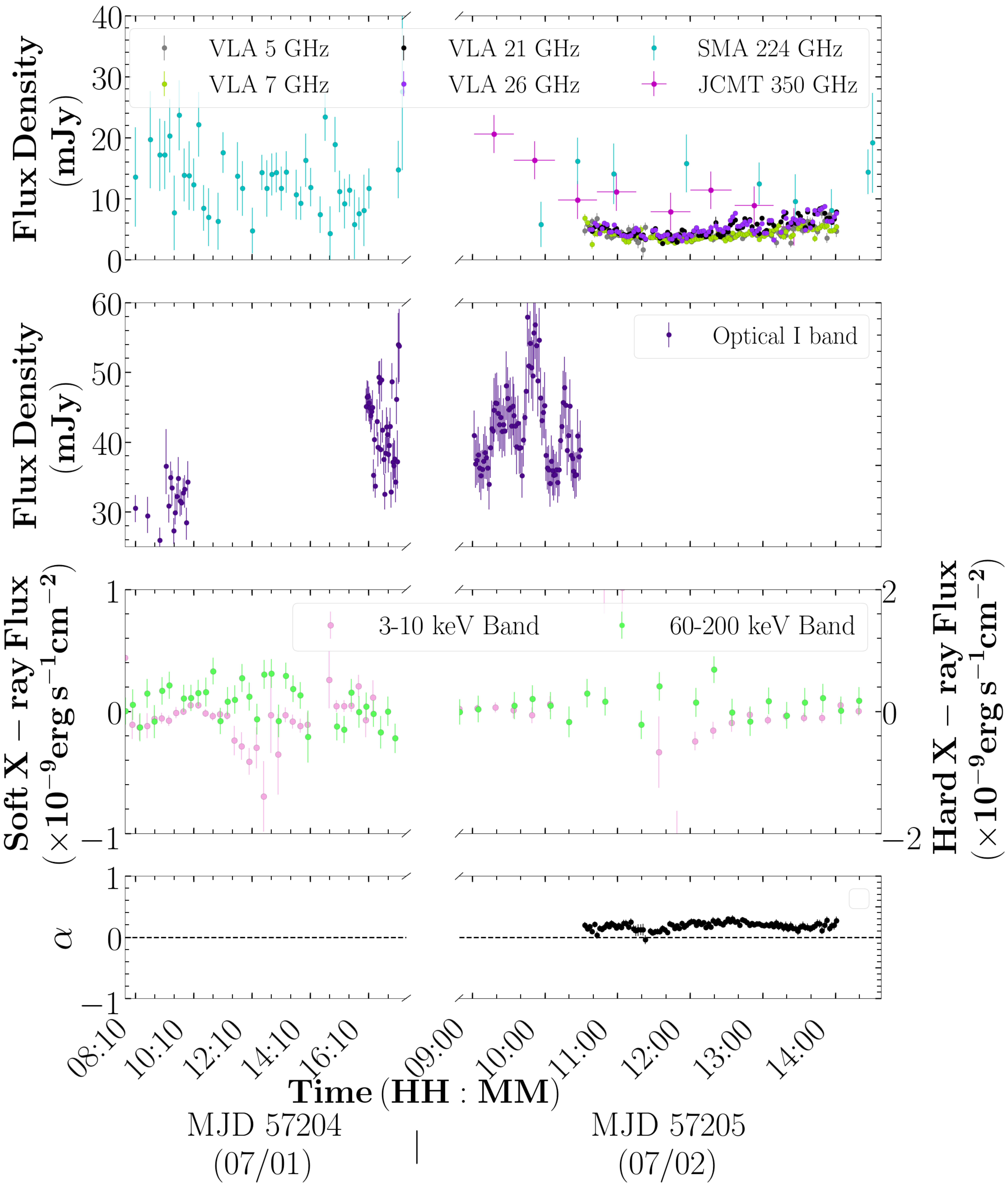}
 \caption{\label{fig:trlc_jul0102}  Time-resolved light curves and spectral indices of V404 Cyg on July 01 and 02 (MJDs 57204 and 57205). The data shown have varying time-bin sizes; 5-26 GHz (2 min), 224 GHz (10/30 min on July 1/2), 350 GHz (30 min), Optical I band (100/75 sec on July 01/02; AAVSO/ \citealt{kim16}), INTEGRAL 3-10 keV (15 min), INTEGRAL 60-200 keV (15 min). Here we have combined the two SMA sidebands, in order to reduce the noise in our short timescale light curves. The horizontal error bars on the JCMT measurements represent the range of the 30 minute SCUBA-2 scans. The dotted line in the bottom panel indicates a spectral index of $\alpha=0$. While no structured flaring activity is observed, the sub-mm emission remains highly variable on July 01, before becoming much more stable a day later.
}
\end{center}
\end{figure}

\begin{figure}
\begin{center}
 \includegraphics[width=1\columnwidth]{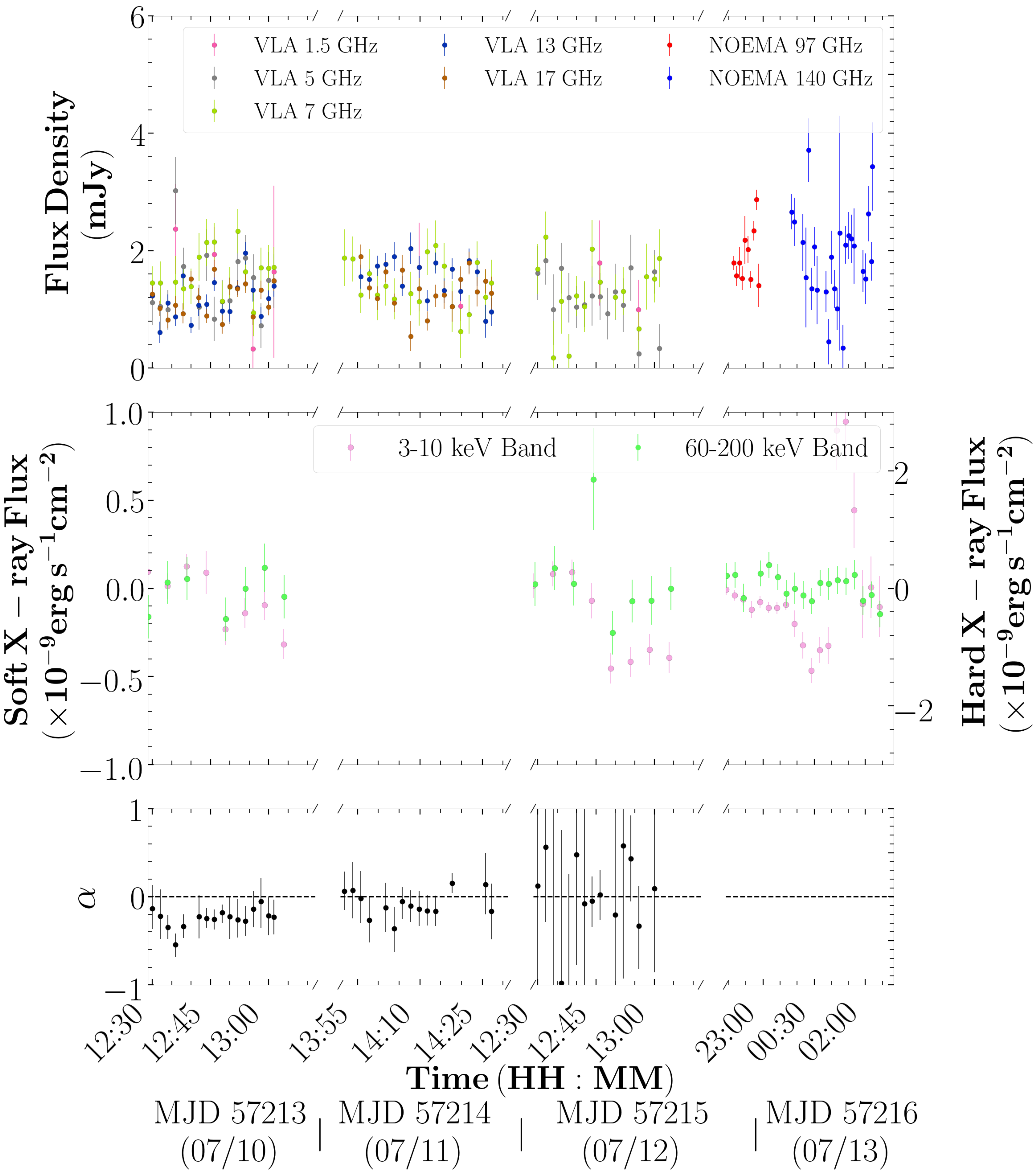}
 \caption{\label{fig:trlc_jul1013}  Time-resolved light curves and spectral indices of V404 Cyg between July 10 and 13 (MJD 57213--57216). The data shown have varying time-bin sizes; 1.5 GHz (10 min), 5-17 GHz (2 min), 97-140 GHz (5 min), INTEGRAL 3-10 keV (5/15 min on MJD 57213-57215/57216), INTEGRAL 60-200 keV (5/15 min on MJD 57213-57215/57216). The dotted line in the bottom panel indicates a spectral index of $\alpha=0$. The radio through sub-mm flux densities are much more constant in these epochs, when compared to our observations one week earlier (see Figure~\ref{fig:trlc_jul0102}).
}
\end{center}
\end{figure}

As V404 Cyg is known to be significantly variable regardless of its brightness, these day-timescale light curves and spectral indices will only display the overall average trend in the data. Therefore, we opted to also search for intra-observation variability in our data. To do this, we created time-resolved light curves (and simultaneous spectral index measurements) of all of our radio through sub-mm observations. These time-resolved light curves, along with simultaneous optical and X-ray data (when available), are displayed in Figures~\ref{fig:trlc_jun1617}, \ref{fig:trlc_jun2630}, \ref{fig:trlc_jul0102}, and \ref{fig:trlc_jul1013}. 
To ensure that any short timescale variations we observe from V404 Cyg are dominated by intrinsic variations, and not atmospheric or instrumental effects, we extracted high time resolution measurements from our calibrator sources as well. We find that the majority of our calibrator observations show relatively constant fluxes (variations $<$ 10\% of the average flux density), except for the SMA data taken on MJD 57189 (see discussion below).

To characterize the amplitude of any intra-observation variability and compare between epochs, we use the fractional RMS statistic, 
\begin{equation}
F_{\rm var}=\sqrt{\frac{S^2-\bar{\sigma}_{\rm err}^2}{\bar{x}^2}}
\end{equation}
where $\bar{x}$ represents the weighted mean of the flux measurements, the sample variance $S^2=\frac{1}{N-1}\sum_{i=1}^{N}(x_i-\bar{x})^2$, and the mean square measurement error $\bar{\sigma}_{\rm err}^2=\frac{1}{N}\sum_{i=1}^{N}\sigma_{\rm err,i}^2$ \citep{akr96,vau03,sad06}. For this paper, we consider $F_{\rm var}< 20$\% as not significantly variable, $20\%<F_{\rm var}< 50$\% as mildly variable, and $F_{\rm var}> 50$\% as highly variable.

In our first epoch on MJD\,57189, the sub-mm emission is highly variable over the short $\sim1.5$ hour observation, with a $F_{\rm var}=71.0\pm1.4\%$ at 224 GHz. Although, we note that this variability appears to be very stochastic (especially when compared to the next epoch taken $\sim 24$ hours later), rather than showing smooth or structured variations that we might expect to see from this high frequency emission probing close to the jet base (e.g., see \citealt{tetarenkoa17}). Upon examining the calibrator light curve for this observation, we noticed that the calibrator source shows an atypically high level of variability within the first half of the observation (i.e., prior to $\sim $18:00 UT), likely due in large part to the very low elevations of these observations (as low as $15$ degrees). Therefore, all of the V404 Cyg variability observed in this epoch may not be intrinsic to the source. Less than a day later, the variability amplitude decreases, along with the average flux level, to just above mildly variable at $F_{\rm var}=51.9\pm0.8\%$ at 224 GHz. While we are only able to sample the sub-mm spectral index in two time bins on MJD\,57190, both show a steep spectral index in this epoch, consistent with that observed from the daily average data (see Figures~\ref{fig:dailylc} \& \ref{fig:trlc_jun1617}).

Our next epoch, taken approximately a week later (MJD\,57195), displayed large scale, structured flaring activity, with lower frequency emission appearing as a smoothed and delayed version of high frequency emission. We have shown that this emission can be well modelled by a series of bi-polar, adiabatically expanding jet ejections (details of this data set are reported in \citealt{tetarenkoa17}). This structured flaring activity (tracing repeated jet ejection events) likely continued intermittently up to MJD\,57199, where we sample a final, large radio flare (peaking at $\sim 200$ mJy at 16 GHz), coinciding with rapid flaring activity at optical and X-ray frequencies (see Figure~\ref{fig:trlc_jun2630}). Following this large flare, the radio through optical emission remains highly variable for another 1--2 days ($F_{\rm var}=71.6\pm0.2\%$ at 140 GHz on MJD\,57200), displaying multiple smaller amplitude flaring events, before the flaring activity ceases, and the variability amplitude drops to mildly variable ($F_{\rm var}=39.0\pm0.2\%$ at 140 GHz on MJD\,57202/57203). The spectral indices during the flaring activity between MJD\,57199--57201 oscillate between steep and inverted on hourly timescales, consistent with the evolving optical depth of adiabatically expanding jet ejecta. After the flaring activity had ceased, the spectral index is much more stable over time, and close to flat.

In the following two days, the sub-mm variability amplitude once again increases to highly variable with $F_{\rm var}=85\pm5\%$ at 224 GHz on MJD\,57204,  before declining to the point where any variance in the data is much less than the measurement errors on MJD\,57205 (see Figure~\ref{fig:trlc_jul0102}). The optical and X-ray emission at this time are not significantly variable ($F_{\rm var}< 10\%$), while the spectral index is stable, and remains inverted across radio through sub-mm bands. 
A week later (MJD\,57213), both the radio and sub-mm flux has dropped by another order of magnitude, while the X-ray flux has remained relatively constant (close to zero) since MJD\,57201 (see Figure~\ref{fig:trlc_jul1013}). All the radio bands are not significantly variable, displaying $F_{\rm var}< 20\%$ (similar to that observed in later radio observations taken in late July and early August; \citealt{plot16}), and the sub-mm variability amplitude is similar to that seen on MJD\,57202/57203. The spectral indices remain flat to slightly inverted during the MJD\,57213--57216 period.

\subsection{Modelling the Flaring Activity}
\label{sec:vdl_jet}
{Out of all the time resolved light curves shown in Figures~\ref{fig:trlc_jun1617} -- \ref{fig:trlc_jul1013}, the radio and sub-mm light curves between MJD\,57199 and 57201 show a distinct morphology. During these days we observe rapid, multi-frequency flaring activity (with the lower frequency emission appearing to be a smoothed, delayed version of the higher frequency emission; see Figure~\ref{fig:trlc_jun2630})}. Given the striking similarity between these data and our earlier work on multi-frequency flaring activity from MJD\,57195 \citep{tetarenkoa17}, we opted to apply the jet model we developed for that data set to the MJD\,57199--57201 light curves. 

\begin{figure*}
   \centering
   \includegraphics[height=8.cm]{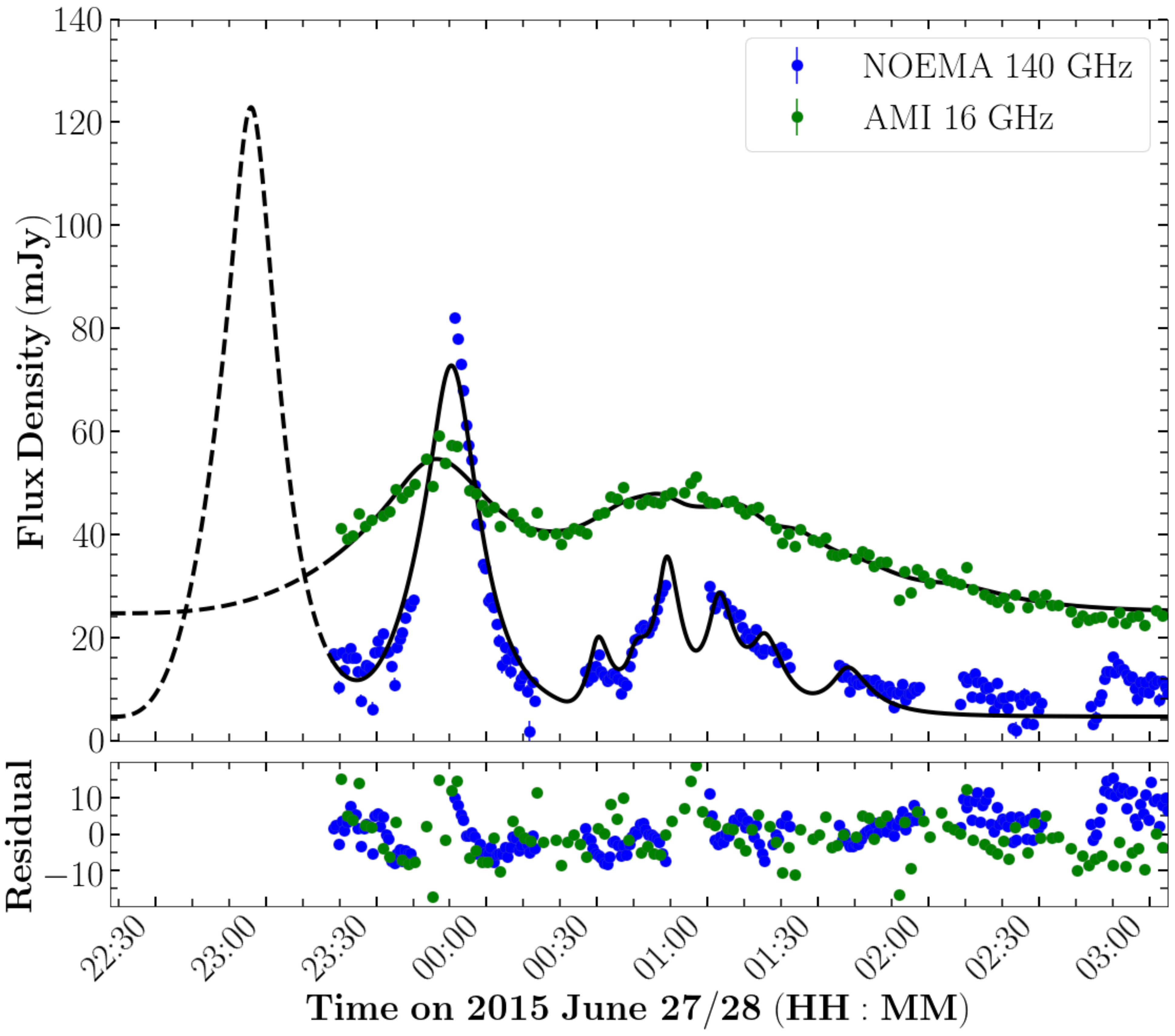} \\
 \caption{\label{fig:vdl}  Radio (AMI 16 GHz) and mm/sub-mm (NOEMA 140 GHz) light curves of V404 Cyg on 2015 
 June 27/28 (MJD 57200/57201). In the {\sl top} panel, we have overlaid our predicted best fit jet model at each frequency on the light curves (black lines, where our model contains contributions from both approaching and receding components for each ejection event). The residuals are shown in the {\sl bottom} panel, where residual=(data--model)/(observational errors). Our best fit model, which contains a total of 5 bi-polar ejection events, can reproduce the overall morphology and flux levels of the emission we observe from V404 Cyg, indicating that this flaring emission is consistent with emission originating from multiple, discrete jet ejection events. Note that we do not attempt to model all of the peaks and wiggles in the mm/sub-mm emission past 02:00 UT, as we do not have the radio frequency coverage to constrain any additional components in the model past this point. See Figure~\ref{fig:vdl_update} for a version of this figure where we decompose the full model into individual approaching and receding components.
 }
 \end{figure*}

{While a detailed description of our jet model is provided in \S 4.2 of \cite{tetarenkoa17}, we provide a brief summary here.
Our V404 Cyg jet model reproduces emission from multiple, discrete, ballistically moving jet ejection events, on top of a constant compact jet component with a power-law spectrum. 
Each of the ejection events consists of the simultaneous launching of identical bi-polar plasma clouds, both of which evolve under the van der Laan \citep{vdl66} synchrotron bubble formalism. 
Additionally, our model folds in both projection and relativistic effects (e.g., relativistic beaming, geometric time delays) for each ejection event.

Here we use the same modelling process detailed in \S 4.3 of \citealt{tetarenkoa17}, where we implement a Markov-Chain Monte Carlo (MCMC) algorithm to fit our light curves on MJD 57200/57201 \citep{for2013}. 
The best fit parameters and their uncertainties\footnote{The uncertainties reported in Table~\ref{table:vdl_table} are purely statistical, only representing confidence intervals on our parameters under the assumption that our model completely represents the data. However, given the residuals with respect to our best-fit model, it is possible that there are physical/instrumental effects in the data that cannot be reproduced by our model. See \S 4.3 of \cite{tetarenkoa17} for a more detailed discussion on this point.} for this fit are shown in Table~\ref{table:vdl_table}, and Figure~\ref{fig:vdl} displays the best fit model overlaid on our light curves (see also Figure~\ref{fig:vdl_update}).} With our best fit model, we find that a total of 5 bi-polar ejection events can reproduce the overall morphology and flux densities of the emission we observe in the June 27/28 (MJD 57200/57201) epoch. {Further, our modelling suggests that the inclination angle of the jet axis changes by up to $\sim 40$ degrees during this series of ejections, which is consistent with the magnitude of jet axis precession independently estimated from a series of resolved jet ejecta observed in an earlier epoch with the VLBA (Miller-Jones et al., Nature Submitted).} Therefore, overall our modelling shows that the flaring emission observed on June 27/28 (MJD 57200/57201) is consistent with emission originating from multiple, discrete jet ejection events. While we also observe multi-frequency flaring in the 2015 June 26/27 (MJD 57199/57200) epoch, there is limited overlap between the mm/sub-mm and radio observations, which makes it difficult to reliably fit this data set with our model. 

{We note that it is possible that the best-fit parameters presented in Table~\ref{table:vdl_table} do not represent a completely unique solution, due to degeneracies in our model (where different combinations of parameters can reproduce similar flaring profiles). Further, the modelling presented in this paper is not as well constrained as our previous application of the model to the flaring from an earlier epoch (MJD 57195; \citealt{tetarenkoa17}), as we only have light curves at 2 simultaneous bands (as opposed to 8 simultaneous bands on MJD 57195) to constrain the model. We caution that the reader should keep this caveat in mind in the further discussion of jet properties in this paper.}

\renewcommand\tabcolsep{2.8pt}
\begin{table*}
\footnotesize
\caption{V404 Cyg Jet Model Best Fit Parameters}\quad
\centering
\begin{tabular}{cccccccccc}
\hline\hline
\multicolumn{10}{c}{{\bf Baseline Flux Parameters$^a$}}\\[0.15cm]
{Epoch}&{${F_{0,\rm cj}}$ (mJy)}&{${\alpha}$}\\[0.25cm]\hline
June 27/28&$ 4.59^{+ 0.06 }_{- 0.06 }$&$ -0.75 ^{+ 0.004}_{- 0.004}$& \\[0.15cm]
&&&&&\\
\hline\hline
\multicolumn{10}{c}{{\bf Individual Jet Ejecta Parameters}$^b$}\\[0.15cm]
{Ejection \,} & 
{${t_{\rm ej}}\,$(HH:MM:SS.S)}&
   {${t_{\rm ej}}\,$(MJD)} &
   {${i\,}$(degrees)}& 
   {$\phi_{\rm obs}$(degrees)}&
   {${\tau_0\,}$}&
   {$p$$^c$}&
   {${F_0\,}$(mJy)}&
   {${\beta_{\rm b}\,}$ (v/c)} &
   {${\beta_{\rm exp}\,}$ (v/c)$^d$}\\[0.25cm]
   \hline
{{June 27/28}}\\[0.1cm]
1 &06-27\,\,\,$\mbox{\formattime{22}{17}{56}}.8 ^{+ 49.13 }_{- 53.84 }$ &$ 57200.9291 ^{+ 0.0006 }_{- 0.0006 }$ & $ 44.07 ^{+ 1.98 }_{- 1.62 }$ & $ 4.55 ^{+ 0.18 }_{- 0.17 }$ & $ 2.09 ^{+ 0.02 }_{- 0.02 }$ & $ 3.59 ^{+ 0.08 }_{- 0.07 }$ & $ 59.48 ^{+ 0.82 }_{- 0.78 }$ & $ 0.037 ^{+ 0.001 }_{- 0.001 }$ & $ 0.0020 ^{+ 0.0001 }_{- 0.0001 }$\\[0.25cm]
2 &06-27\,\,\,$\mbox{\formattime{23}{14}{19}}.3 ^{+ 12.56 }_{- 11.48 }$ &$ 57200.9683 ^{+ 0.0001 }_{- 0.0001 }$ & $ 53.45 ^{+ 1.49 }_{- 1.62 }$ & $ 5.86 ^{+ 0.20 }_{- 0.16 }$ & $ 1.99 ^{+ 0.01 }_{- 0.01 }$ & $ 3.27 ^{+ 0.04 }_{- 0.04 }$ & $ 34.29 ^{+ 0.20 }_{- 0.20 }$ & $ 0.020 ^{+ 0.001 }_{- 0.001 }$ & $ 0.0017 ^{+ 0.0001 }_{- 0.0001 }$\\[0.25cm]
3 &06-28\,\,\,$\mbox{\formattime{00}{23}{46}}.5 ^{+ 16.64 }_{- 16.66 }$ &$ 57201.0165 ^{+ 0.0002 }_{- 0.0002 }$ & $ 34.56 ^{+ 2.46 }_{- 2.84 }$ & $ 6.47 ^{+ 0.73 }_{- 0.53 }$ & $ 1.50 ^{+ 0.01 }_{- 0.01 }$ & $ 1.99 ^{+ 0.01 }_{- 0.01 }$ & $ 12.60 ^{+ 0.24 }_{- 0.27 }$ & $ 0.050 ^{+ 0.002 }_{- 0.002 }$ & $ 0.0032 ^{+ 0.0004 }_{- 0.0003 }$\\[0.25cm]
4 &06-28\,\,\,$\mbox{\formattime{00}{42}{13}}.7 ^{+ 13.14 }_{- 12.60 }$ &$ 57201.0293 ^{+ 0.0002 }_{- 0.0001 }$ & $ 63.30 ^{+ 1.23 }_{- 1.55 }$ & $ 2.37 ^{+ 0.15 }_{- 0.12 }$ & $ 1.51 ^{+ 0.03 }_{- 0.01 }$ & $ 2.01 ^{+ 0.06 }_{- 0.02 }$ & $ 24.09 ^{+ 0.60 }_{- 0.96 }$ & $ 0.107 ^{+ 0.004 }_{- 0.005 }$ & $ 0.0040 ^{+ 0.0004 }_{- 0.0003 }$\\[0.25cm]
5 &\phantom{0}06-28\,\,\,$\mbox{\formattime{01}{08}{23}}.2 ^{+ 62.68 }_{- 133.28 }$ &$ 57201.0475 ^{+ 0.0007 }_{- 0.0015 }$ & $ 19.54 ^{+ 1.06 }_{- 1.10 }$ & $ 4.74 ^{+ 0.29 }_{- 0.25 }$ & $ 1.56 ^{+ 0.04 }_{- 0.04 }$ & $ 2.12 ^{+ 0.09 }_{- 0.10 }$ & $ 10.23 ^{+ 0.52 }_{- 0.30 }$ & $ 0.067 ^{+ 0.003 }_{- 0.002 }$ & $ 0.0018 ^{+ 0.0002 }_{- 0.0001 }$\\[0.25cm]
\hline
\end{tabular}\\
\begin{flushleft}
\footnotesize
{$^a$ {The baseline flux (representing contributions from a compact jet) is best fit by a single power-law, with amplitude ${F_{0,\rm cj}}$ at 140 GHz, and spectral index $\alpha$.}}\\
{$^b$ Ejecta parameters are defined as follows: ejection time (${t_{\rm ej}}$), inclination angle of the jet axis ($i$), jet opening angle ($\phi_{\rm obs}$), synchrotron optical depth (at 140 GHz) at the time of peak flux density ($\tau_0$), peak flux density of the ejecta component ($F_0$), and the bulk ejecta speed ($\beta_b$).}\\
{$^c$ The index of the electron energy distribution, $p$, is not a fitted parameter but rather is solved for using values of $\tau_0$. Similar to the situation discussed in our earlier modelling work \citep{tetarenkoa17}, the large range of energy indices ($p$) for the ejecta found here is not entirely physical for a single source. We believe that more extreme values of the energy index could be mimicking the effect of physics that has not been included in our model.}\\
{$^d$The expansion velocity, $\beta_{\rm exp}$, is not a fitted parameter but rather is solved for using values of $\beta_{\rm b}$, $i$, $\phi_{\rm obs}$.}\\
\end{flushleft}
\label{table:vdl_table}
\end{table*}
\renewcommand\tabcolsep{6pt}

\subsection{Cross-correlation analysis}
\label{sec:ccf}
To search for time lags between different frequency bands, {we computed cross-correlation functions (CCFs) of our time-resolved light curves in all epochs}, using the z-transformed discrete correlation function (ZDCF; \citealt{alex97,alex13a}). We chose to use the ZDCF algorithm, as this method has been shown to provide a more robust estimate of the CCFs for sparse, unevenly sampled light curves, when compared to the classic discrete correlation function (DCF; \citealt{ed88}) or the interpolation method \citep{gask87}. To obtain an estimate of the CCF peak (indicating the strongest positive correlation, and thus the best estimate of any time-lag between the light curves from different frequency bands), with corresponding uncertainties, we utilize the maximum likelihood method\footnote{This method estimates a fiducial interval rather than the traditional confidence interval. The approach taken here is similar to Bayesian statistics, where the normalized likelihood function (fiducial distribution) is interpreted as expressing the degree of belief in the estimated parameter, and the 68\% interval around the likelihood function's maximum represents the fiducial interval (68\% of the likelihood-weighted ensemble of all possible CCFs reach their peaks within this interval).} described in \citealt{alex13a}. {Additionally, to estimate the significance level of any peak in the CCF, we perform a set of simulations.  For these simulations, we randomize each radio light curve 1000 times (i.e., Fourier transform the light curves, randomize the phases, then inverse Fourier transform back, to create simulated light curves that share the same power spectra as the real light curves), and calculate the CCF for each randomized case}. We then determine the fraction of simulated CCF data points above the peak CCF level in the original CCF run.
Performing these simulations allows us to quantify the probability of false detections in our CCFs, by accounting for stochastic fluctuations and intrinsic, uncorrelated variability within each radio light curve.
\begin{figure*}
   \centering
   \subfloat{\includegraphics[height=7.5cm]{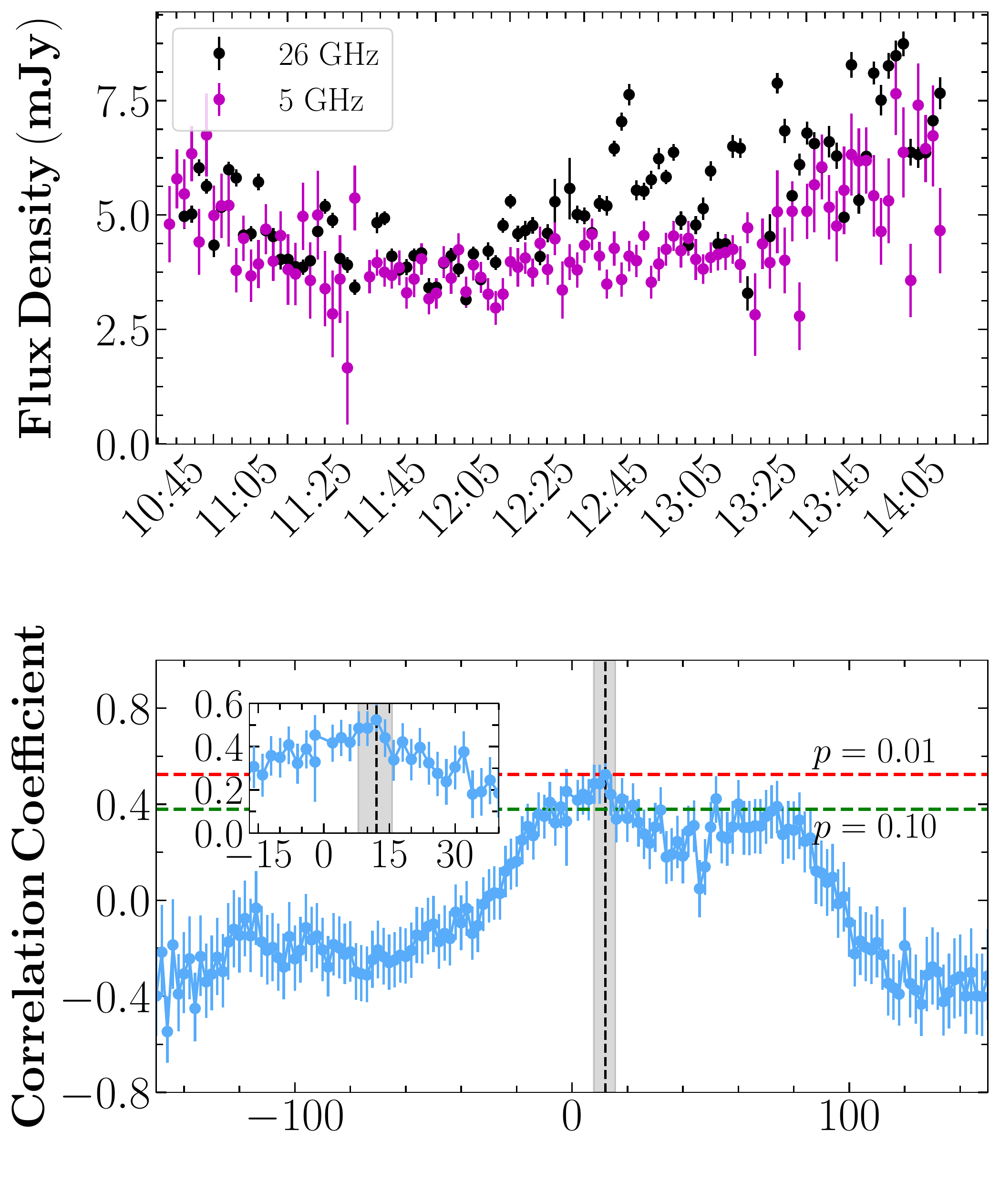}}
    \subfloat{\includegraphics[height=7.5cm]{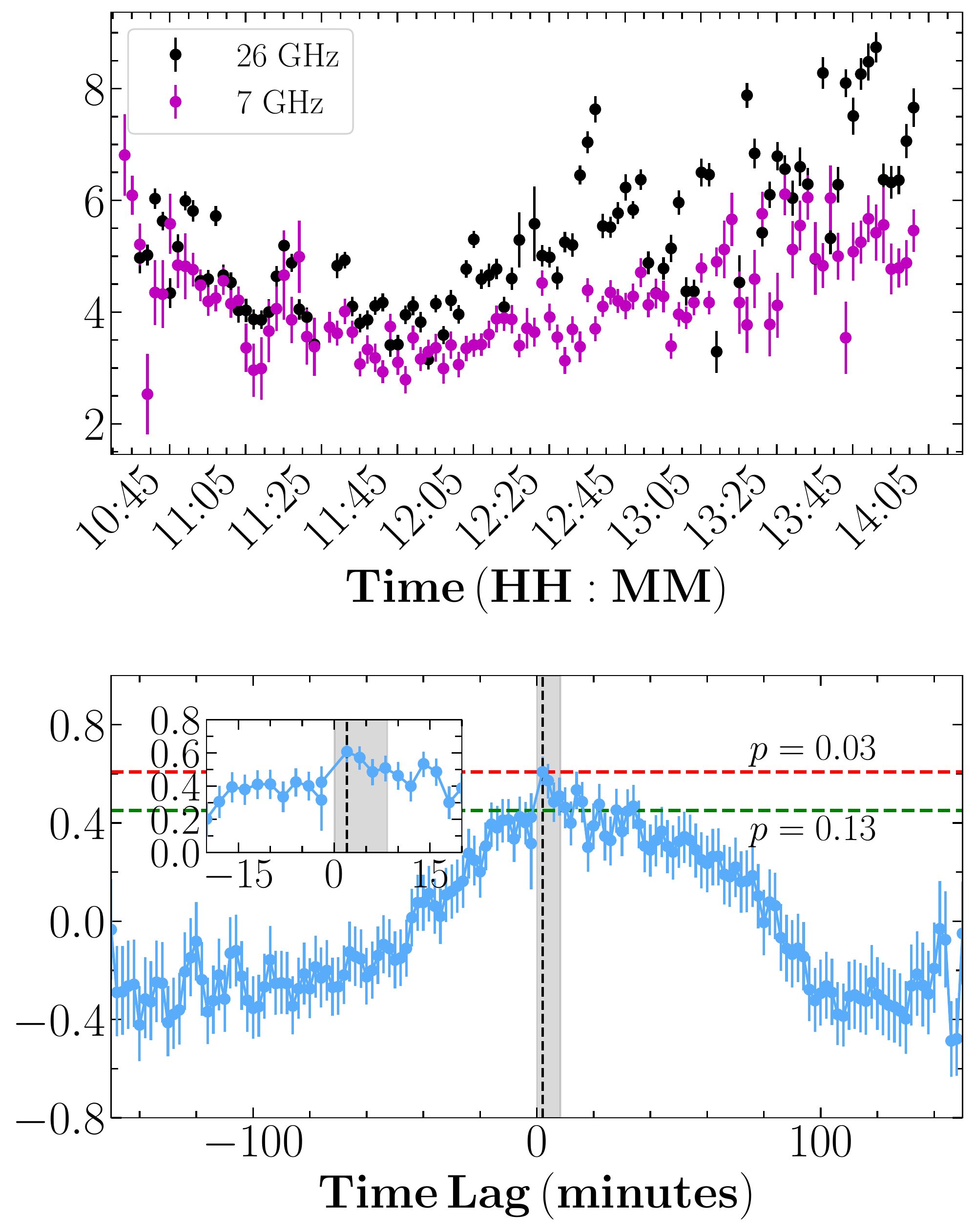}}
   \subfloat{ \includegraphics[height=7.5cm]{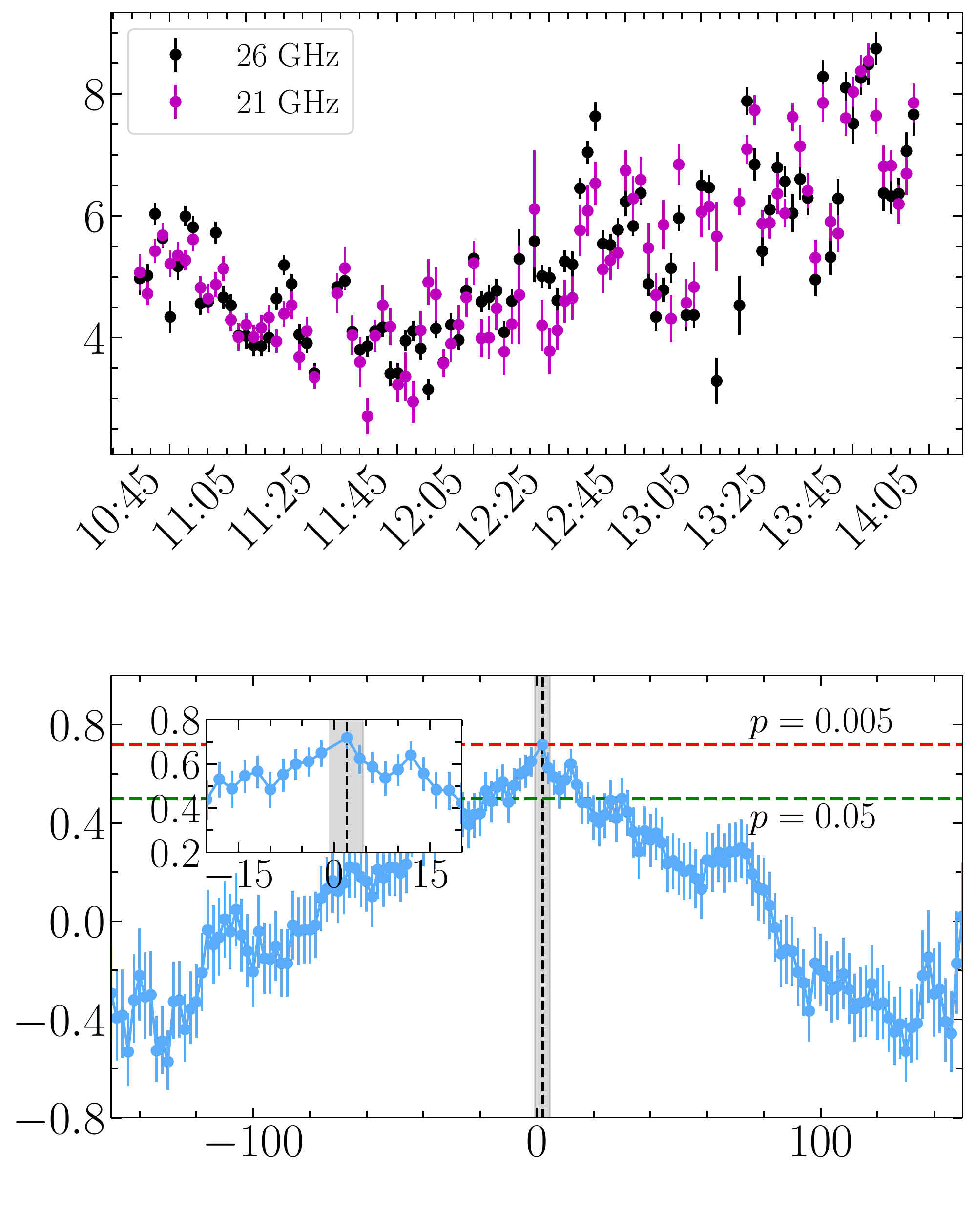}} \\
 \caption{\label{fig:lag}  Radio light curves (zoomed in versions of Figure~\ref{fig:trlc_jul0102}; {\sl top panels}) and cross-correlation functions (CCFs; {\sl bottom panels}) between the VLA radio frequency bands on July 02 (MJD 57205). The peak of each CCF (indicating the strongest positive correlation) is shown by the black dotted line, with the 68\% fiducial confidence interval indicated by the shaded grey region, and different significance levels indicated by the red and green dotted lines (see \S\ref{sec:ccf} for details). The insets in each panel display a zoomed view of the region surrounding the peak of each CCF. A positive time lag indicates that the lower frequency band lags behind the higher frequency band. We observe a clear time lag between the 26 GHz and 5 GHz bands ({\sl bottom left}; $12.0^{+3.7}_{-4.2}$ min), however, our measured lags between the 26 GHz and the 7 GHz ({\sl bottom middle}) / 21 GHz ({\sl bottom right}) bands are consistent with zero lag. 
 }
 \end{figure*}

{While we performed the above CCF analysis for all of the radio through X-ray data sets for which we had overlapping, time-resolved data, we only find clear evidence of time lags during the July 02 (MJD 57205) epoch (see Figure~\ref{fig:lag}). In particular, we measure a time lag between the 26 GHz radio band and the 5 GHz radio band of $12.0^{+3.7}_{-4.2}$ minutes.} However, the measured time lags between the 26 GHz band and the 7/21 GHz bands are consistent with a zero lag within the uncertainty limits (where the 1 $\sigma$ upper limits for the 26 GHz to 7/21 GHz lags are $<10$ min and $<5$ min, respectively). Further, the optical I band flaring activity observed in this epoch (see Figure~\ref{fig:trlc_jul0102}) is unlikely to be correlated with this radio emission, as a simple jet model ($z_0\propto 1/\nu$, where $z_0$ represents the distance down the jet axis from the black hole) paired with our our detected radio lag predicts a $\sim 15$ min lag between the I band and 5 GHz, rather than the hours between the optical and radio flaring observed in the light curves (see Figures~\ref{fig:trlc_jul0102} and \ref{fig:lag}). However, as the radio and I band light curves do not overlap in this epoch, we can not rule out a correlation between the two.
Therefore, we are unable to conclusively determine if a trend with frequency, where the lower frequency bands always lag the higher frequency bands (and the lag increases as the frequency decreases in the comparison band), exists in our CCFs. Such a trend is expected from emission originating in a compact jet, as lower frequency emission is expected to originate from a region further down the jet axis from the black hole, and these time lags between radio bands could trace the propagation of material downstream along the jet \citep{mal03,cas10,gan17,vinc18}.

{As the jet model we used in \S\ref{sec:vdl_jet} can predict lags between different frequency bands for each jet ejection event, it is of interest to compare the model predicted lags from data on MJD 57200/57201 and the CCF predicted lags from data on MJD 57205. In particular, our best-fit jet model from MJD 57200/57201 predicts time lags between 26 and 5 GHz of 24--59 min, between 26 and 7 GHz of 17--44 min, and between 26 and 21 GHz of 2--5 min, for different ejection events. Therefore, the jet model predicted lags between 26 and 5 GHz are all larger than our CCF measured lag. These differing lags likely indicate varying jet properties at different phases of the outburst (i.e., decline from a major flare vs. flat-spectrum compact jet emission at a much lower level), such as bulk speed, inclination angle, opening angle, or electron energy distributions, between the MJD 57200 and 57205 epochs.}

\begin{figure*}
\begin{center}
 \includegraphics[width=1.5\columnwidth]{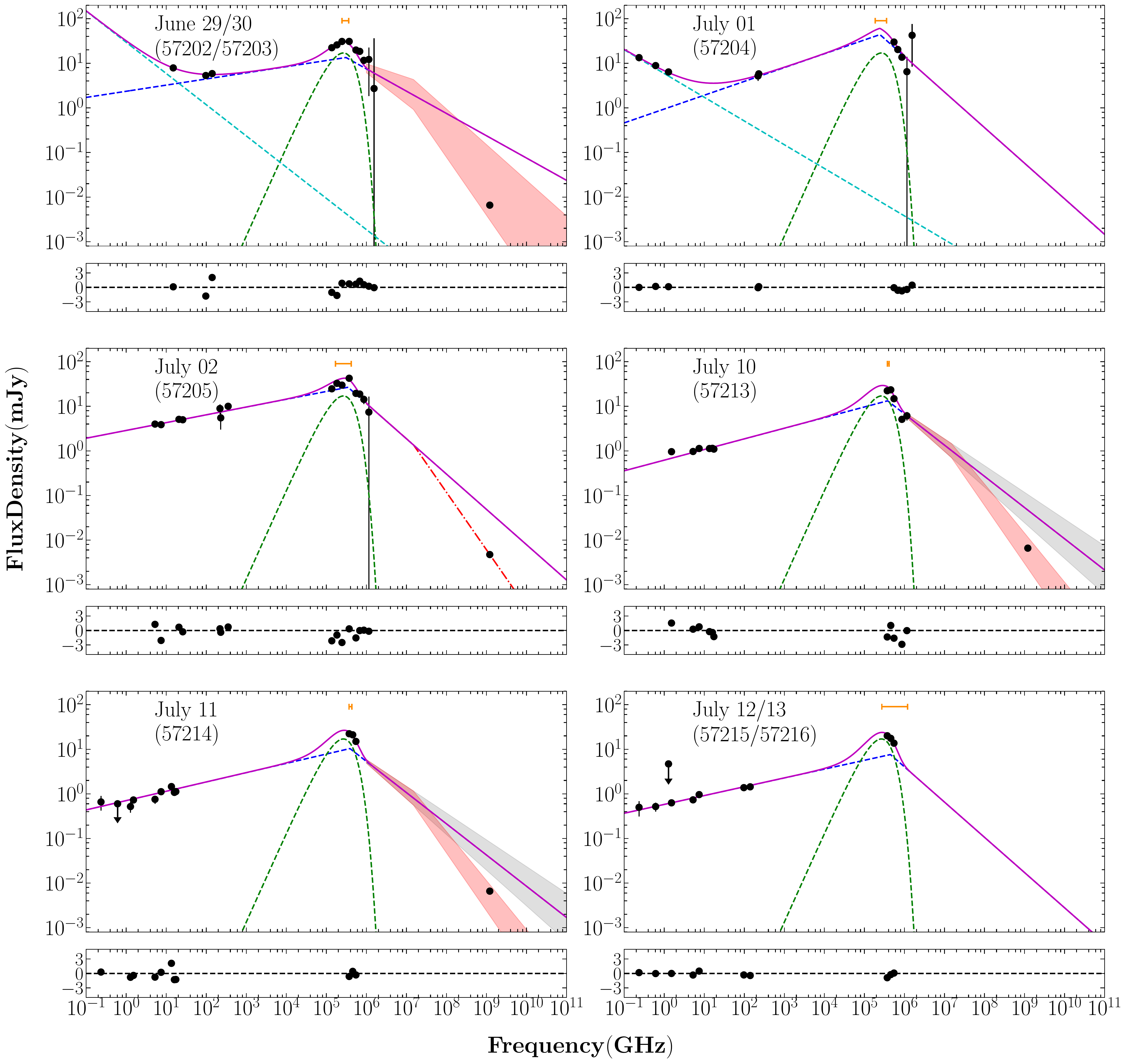}
 \caption{\label{fig:seds}  Broad-band spectra of V404 Cyg during the decay of the 2015 outburst. The {\sl top} panel in each broad-band spectrum displays the broad-band photometric data (black markers), and the best-fit model (where the {\em Swift XRT}/{\em Chandra} X-ray data points are not included in these fits), in each epoch. The solid purple line represents the total model, and the dotted lines represent the model components (green is the companion star component, dark blue is the compact jet component, and cyan is the fading jet ejecta component). The orange bars/arrows represent constraints on the location of the spectral break. The dash-dotted red line and shaded red/gray regions represent constraints on the synchrotron jet contribution to the X-ray emission (where our jet model over-predicts X-ray flux unless a sychrotron cooling break is considered in the spectrum; see \S\ref{sec:specb} for details). The {\sl bottom} panel of each broad-band spectra represents the residuals, where residual=(data-model)/(observational errors). The emission in these broad-band spectra is consistent with being dominated by emission from a compact jet, with a spectral break in the near-IR and optical frequency bands.
}
\end{center}
\end{figure*}

\renewcommand\tabcolsep{8pt}
 \begin{table*}
 \footnotesize
\caption{Best-fit parameters for broad-band spectral modelling$^{a,b}$}\quad
\centering
\begin{tabular}{ cccccccc }
 \hline\hline
  {Date} &{$\nu_{\rm break}$  (GHz)}&{$F_{\rm break}$ (mJy)}&{ $\alpha_{\rm thick}$ }&{$\alpha_{\rm thin}$ }&{$L_{\rm jet}\,({\rm erg\,s}^{-1})$}&{ $F_{15\,{\rm GHz}}$ (mJy)}&{ $\alpha_{{\rm thin},2}$} \\[0.15cm]
  \hline
June 29/30&$(3.10_{-0.63}^{+0.50})\times10^5$&$13.48_{-0.90}^{+0.53}$&$0.14_{-0.01}^{+0.01}$&$-0.50_{-0.29}^{+0.22}$&$(5.47_{-0.97}^{+1.14})\times10^{35}$&$4.45_{-0.32}^{+0.25}$&$-0.70$\\[0.15cm]
July 01&$(2.44_{-0.57}^{+1.12})\times10^5$&$43.97_{-0.29}^{+0.50}$&$0.31_{-0.03}^{+0.11}$&$-0.80_{-0.10}^{+0.18}$&$(1.42_{-0.28}^{+0.33})\times10^{36}$&$1.33_{-0.69}^{+0.56}$&$-0.53_{-0.16}^{+0.11}$\\[0.15cm]
July 02&$(3.48_{-1.79}^{+0.69})\times10^5$&$26.71_{-4.14}^{+2.10}$&$0.17_{-0.01}^{+0.02}$&$-0.79_{-0.18}^{+0.37}$&$(9.45_{-2.91}^{+2.81})\times10^{35}$&\dots&\dots\\[0.15cm]
July 10&$(3.81^{+0.06}_{-0.31})\times10^{5}$&$13.27^{+0.92}_{-1.56}$&$0.24_{-0.01}^{+0.01}$&$-0.70$&$(5.25^{+0.92}_{-0.81})\times10^{35}$&\dots&\dots\\[0.15cm]
July 11&$(3.89^{+0.44}_{-0.20})\times10^{5}$&$10.38^{+1.59}_{-1.64}$&$0.20_{-0.02}^{+0.02}$&$-0.70$&$(4.22^{+1.10}_{-0.79})\times10^{35}$&\dots&\dots\\[0.15cm]
July 12/13&$(4.48^{+7.58}_{-1.75})\times10^{5}$&$7.62^{+1.21}_{-2.67}$&$0.19_{-0.02}^{+0.04}$&$-0.70$&$(3.03^{+1.19}_{-1.05})\times10^{35}$&\dots&\dots\\[0.15cm]

\hline
\end{tabular}\\
\begin{flushleft}
 \footnotesize
{$^a$ Columns from left to right: spectral break frequency, flux at the spectral break, optically thick spectral index, optically thin spectral index, integrated compact jet power ($L_{\rm jet}=4\pi D^2\int{\nu L_\nu\,d\nu}$, from 1.5--$1.2\times10^{6}$ GHz) given our best-fit model, flux (at 15 GHz) and spectral index of additional optically thin power law component. The radius and temperature of the companion star were fixed in all fits ( $R_\star=5.71R_\odot$, $T_\star=0.784T_\odot$; \citealt{gallo7}).}\\
{$^b$ Note that we fix the optically thin spectral index to a value of $-0.7$ for the July 10, July 11 and July 12/13 epochs, as we do not have sufficient data to place accurate constraints on this parameter in our fitting.}\\
\end{flushleft}
\label{table:sed_tab}
\end{table*}
\renewcommand\tabcolsep{6pt}

\subsection{Broad-band Spectra}
\label{sec:specb}
{In the epochs following the structured flaring activity (MJD\,57203--57216), where the emission is much more constant (showing minimal flux variability within an observation), we constructed broad-band spectra} to track the spectral evolution of the jet emission as V404 Cyg decayed towards quiescence {(including additional radio data from the Giant Metrewave Radio Telescope reported in \citealt{chpoo17}, to add lower frequency coverage at 0.235, 0.610, and 1.280 GHz)}. We fit these radio through optical/UV broad-band spectra with a phenomenological multi-component model, consisting of a broken power law (representing compact jet emission), {a black-body ($R_\star=5.71R_\odot$, $T_\star=0.784T_\odot$, representing the known companion star; \citealt{gallo7})}, and in two epochs, an additional single power law (representing emission from fading jet ejecta).
To fit these spectra we use a MCMC algorithm \citep{for2013}, where the best fit result is taken as the median of the one-dimensional posterior distributions, and the uncertainties are reported as the range between the median and the 15th percentile (-), and the 85th percentile and the median (+), corresponding approximately to $1\sigma$ errors.
These broad-band spectra are displayed in Figure~\ref{fig:seds}, and the best-fit model parameters are reported in Table~\ref{table:sed_tab}. {We note that while accretion disc emission has been known to contribute to the optical/UV emission in broad-band spectra of BHXBs (\citealt{khat10} estimate $<3\%$ accretion disc contamination during quiescence for V404 Cyg, although the accretion disc is much brighter in outburst than in quiescence; \citealt{bern16}), we do not include an accretion disc component in our model presented here (e.g., \citealt{hynes2002,hyn05}). While we could reasonably reproduce the optical/UV emission in our broad-band spectra with a cool ($T\sim 3000$ K), highly truncated ($R\sim 10^4\, R_g$), viscous disc (where irradiation is not necessary to describe the spectral shape), the integrated flux over this disc emission implies a physically improbable mass transfer rate (e.g., on MJD 57205 $\sim2\times10^{-9}\,M_\odot/{\rm yr}^{-1}$) through the disc for this scenario. Therefore, we favour a model where the jet dominates the optical/UV emission in our broad-band spectra.}

The broad-band spectra constructed from data on MJD\,57202--57205, are well fit by a broken power-law, where we detect the spectral break in the near-IR bands. The OIR emission from the companion star is fainter than the jet emission on MJDs 57204 and 57205, contributing little to the overall broad-band spectra (although the jet and companion star show similar flux levels in the OIR on MJD 57202/57203). {However, on MJDs\,57202/57203, and 57204, we require an additional power-law component, to account for the excess emission at radio frequencies}\footnote{{Note that we have fixed the spectral index of this jet ejecta component to a typical value of $-0.7$ in the MJD 57202/57203 epoch, as we only have a single radio data point to constrain this component. However, as we have three radio data points in the MJD 57204 epoch, we allow this parameter to vary in the fit for that epoch.}}. The broken power-law emission is characteristic of a compact jet, while the additional power-law component could originate in emission from fading jet ejecta, potentially launched during the flaring period 2--3 days prior to these epochs. 
Further, while we see very little evolution in the location of the spectral break across these three broad-band spectra, the optically thin spectral index may steepen over time, while the optically thick spectral index may flatten over time (although, given the large uncertainties in some epochs, it is difficult to determine if we see an evolutionary trend in this spectral index; e.g., MJD 57204, where we only sample the optically thick part of the spectrum in two closely placed sub-mm bands).

The broad-band spectra constructed from data on MJD\,57213--57216, are also well fit by a broken power-law, where we detect the spectral break at frequencies as high as the optical bands. This indicates that the spectral break has moved to higher frequencies over the $\sim 1$ week timescale between these epochs and the previous three epochs. Further, during these later epochs, the jet emission has faded by $\sim$ an order of magnitude, and thus the emission from the companion star contributes much more to the overall spectral shape in these broad-band spectra.

{We note that all of the above conclusions are dependent upon the assumption that other emission sources (e.g., accretion disc emission, irradiation of the companion star by X-rays), are not significantly contaminating the OIR/UV bands in our broad-band spectra. For instance, X-rays emitted during the final major flaring event of the outburst (occurring a few days prior to the spectra presented in this work) 
could have potentially caused the emission from the companion star to be much brighter than normal (through the irradiation process), and in turn contribute more to the OIR/UV part of the broad-band spectrum. In this case, we would expect a smaller jet contribution to the OIR/UV emission. In fact, this scenario may explain the larger deviations between the radio/sub-mm data and our best-fit model on MJD 57213. The presence of a hotter companion star component, producing more OIR/UV flux, would allow for a flatter jet spectrum at lower frequencies, that would be more representative of the radio/sub-mm data in this epoch.}

Further, given the high spectral break frequency measured here, it is of interest to explore whether the synchrotron jet emission could be dominating the emission in the X-ray bands during our sampled epochs. To test this scenario, we have also included the available quasi-simultaneous {\em Swift XRT}/{\em Chandra} X-ray flux measurements (in the 0.5--10 keV band) within our broad-band spectra (Figure~\ref{fig:seds} and Table~\ref{table:other_flux}). Simply extrapolating the optically thin part of the jet spectrum to the X-ray bands in the X-ray sampled epochs (blue dotted lines on Jun 29/30 \& July 02, and gray shading representing a range of expected optically thin spectral indexes between $\alpha=-0.6$ and $\alpha=-0.8$ on July 10 \& 11) clearly over-predicts the X-ray flux. Therefore, we consider the possibility where a second break, representing a synchrotron cooling break\footnote{We note that to the best of our knowledge the synchrotron cooling break has only been detected in the broad-band spectra of one BHXB so far (MAXI J1836--194), where the cooling break was found in the optical bands between $(3-4.5)\times10^{14}$ Hz
\citep{rus14}.} (due to the highest-energy electrons losing their energy through radiation on timescales faster than the dynamical time scale; \citealt{sari98,pm12,rus12,rus14}), occurs between the UV and X-ray bands in the synchrotron spectrum. To place constraints on the location of the cooling break in this case, we consider the July 02 (MJD 57205) epoch, as this is the only epoch in which we have data sampling the optically thin part of the jet spectrum, no contribution from jet ejecta, and an X-ray measurement. Through refitting the July 02 spectrum, including the X-ray data point, adding a cooling break (where the spectral index after the cooling break is steeper by $\Delta\alpha=0.5$; \citealt{sari98,rus14}) in the model, and keeping all other parameters fixed at the original best-fit values, we find $\nu_{\rm coolbr}=(1.5^{+0.7}_{-0.5})\times10^{16}$ Hz (where the cooling break version of the model is displayed as a red dot-dashed line in the middle-left panel of Figure~\ref{fig:seds}). Given this cooling break measurement, and estimates of the optically thin spectral index (we assume spectral indexes between $\alpha=-0.6$ and $\alpha=-0.8$ in epochs where we have no constraint on this parameter), we also extrapolate the synchrotron spectrum to the X-ray bands in the June 29/30, as well as the July 10 \& 11 epochs (displayed as red shading in the top-left, middle-right, and bottom-left panels of Figure~\ref{fig:seds}). Overall, we find that the jet synchrotron emission could reasonably be producing a significant fraction of the X-ray flux in these epochs.
Further discussion of the plausibility of this scenario is presented in \S\ref{sec:cjdisc}.

\section{Discussion} 
\label{sec:dis}

Throughout the June 2015 outburst of V404 Cyg, the jet emission we observe displays a wide range of intensities (spanning over 3 orders of magnitude between the brightest and faintest epochs), and the spectral and variability properties of the jet emission change dramatically throughout the outburst (on timescales of minutes to days). In this work, we have presented detailed diagnostics of this jet emission, and in the following sections we discuss jet properties and evolution in V404 Cyg, as well as draw comparisons to the jet emission observed in the 1989 outburst, and the December 2015/January 2016 mini-outburst.

\subsection{Jet ejecta behaviour}
For the first $\sim13$ days of the June 2015 outburst, the jet emission from V404 Cyg appears to be dominated by emission from discrete jet ejections, as evidenced by the structured multi-frequency flaring activity in the light curves, and the rapidly oscillating radio through sub-mm spectral indices (consistent with the evolving optical depth of these expanding ejecta; see Figure~\ref{fig:dailylc} \& \ref{fig:trlc_jun2630}).
In recent work \citep{tetarenkoa17}, we developed a jet ejecta model for V404 Cyg that could reproduce the brightest multi-frequency flaring emission detected during the outburst (on MJD 57195), and in turn allow us to probe jet speeds, energetics, and geometry. To examine how the jet ejecta properties could evolve throughout the outburst,  in this paper we have presented model fits to another multi-frequency flaring data set, occurring 5 days following the brightest epoch. We find that these later multi-frequency flaring episodes can also be well represented by emission from a series of jet ejections (Figure~\ref{fig:vdl}). Upon comparing the jet ejecta properties between our modelled data sets, the later epochs tend to show fainter ejecta (tens to hundreds of mJy, rather than thousands of mJy), with lower bulk speeds ($<0.1c$ versus $0.2-0.6c$ on MJD 57195; \citealt{tetarenkoa17}), and longer periods between ejections (on the order of hours, rather than minutes). This suggests that the ejecta properties changed throughout the flaring period, becoming slower, less energetic, and less frequent as the outburst progressed, before the discrete jet ejections stopped all together.

{If the jet ejecta launched from V404 Cyg are powered by the accretion flow, then the radio/sub-mm emission we observe should be correlated with the optical/X-ray emission (where the optical emission could originate in the jet base/acceleration region and/or be reprocessed X-ray emission from the accretion flow; e.g., \citealt{gan17}). In particular, we might expect\footnote{We note that while multi-band flaring activity detected in some epochs may appear to be occurring simultaneously, observing no delay between X-ray/optical/sub-mm/radio emission does not make physical sense in the accretion flow/jet picture. Emission from the X-ray/optical/sub-mm/radio bands are emitted from different regions that are expected to be separated by measurable light travel times. As such, any appearance of simultaneity is more likely to be a sampling effect in our light curves.} to observe optical/X-ray flaring counterparts preceding our radio/sub-mm flares, where the ejection times predicted by our modelling should coincide with the optical/X-ray flaring.
Interestingly, the beginning of the optical flaring complex observed on June 27 at $\sim$ 23:00 UT coincides with the predicted ejection time for the largest mm/sub-mm flare (ejection 2 in Table~\ref{table:vdl_table}, also see Figure~\ref{fig:vdl_ejtimes}) observed in this epoch. Additionally, while we were unable to model the radio/sub-mm flaring on June 26, 
if we assume similar delay timescales ($\sim10-40$ min between mm/sub-mm and optical/X-ray) from our modelled epoch, it seems plausible that the rapid optical/X-ray activity on June 26 could also be related to the radio/sub-mm flaring we observe during this epoch (see Figure~\ref{fig:trlc_jun2630}). However, not all of the ejection events we model on June 27 have clear optical counterparts (e.g., ejections 3, 4 and 5; see Figure~\ref{fig:vdl_ejtimes}). The explanation for the lack of clear optical/X-ray counterparts to some radio/sub-mm flares in this system is not known, but additional factors such as a precessing accretion disc (Miller-Jones et al., Nature Submitted) obscuring the jet base or emission being absorbed by the strong accretion disc wind detected from the inner accretion disc in this source \citep{mun16}, could be affecting the optical/X-ray emission we observe, thus making it more difficult to identify counterparts to the radio/sub-mm flares.
Therefore, our observations provide hints of a possible correlation between radio/sub-mm and optical/X-ray emission during jet ejection events in V404 Cyg, but we do not fully understand the connection between the accretion flow and jet emission in this system.}

\subsection{Compact jet behaviour}
\label{sec:cjdisc}
Following the flaring period, the jet emission from V404 Cyg switches to being dominated by a compact jet, as evidenced by the flat/inverted (radio through sub-mm) spectral indices, and the absence of large-scale flaring activity in the time-resolved light curves in all bands (see Figures~\ref{fig:trlc_jul0102} \& \ref{fig:trlc_jul1013}).

Through applying a phenomenological model to the broad-band emission at this stage in the outburst, we find that the compact jet is the dominant source of emission in V404 Cyg from the radio through optical/UV bands (see Figure~\ref{fig:seds}, where the broad-band spectra can be well fit by a broken power law, characteristic of compact jet emission).
With this modelling, we initially detect the optically thick to thin synchrotron jet spectral break in the near-IR bands ($\sim2-3\times10^{14}$ Hz), and find that the spectral break could reach as high as the optical bands in our final epochs ($4.48\times10^{14}$ Hz). This high spectral break frequency is atypical for BHXBs (typical values for the spectral break in BHXBs are $10^{11-14} {\rm \, Hz}$; \citealt{rus12}), but consistent with data from the
1989 outburst, which also display a high spectral break frequency ($\sim1.8\times10^{14}$ Hz; \citealt{rus12}).
These results are also
in agreement with the recent work of \cite{maitra17}, who present several lines of evidence to suggest that the optical emission was dominated by the compact jet on MJD 57200, and theorize the spectral break lies above the optical V band at this time ($\sim5.5\times10^{14}$ Hz). As the spectral break probes the jet base (where jet particles are first accelerated to high energies), \cite{maitra17} theorize that a spectral break at such a high frequency suggests the jet base was very compact and energetic at this point in the outburst. If this is the case, our spectral break measurements imply these conditions persisted as the system faded toward quiescence. 

\begin{figure*}
\centering
 \includegraphics[width=2\columnwidth]{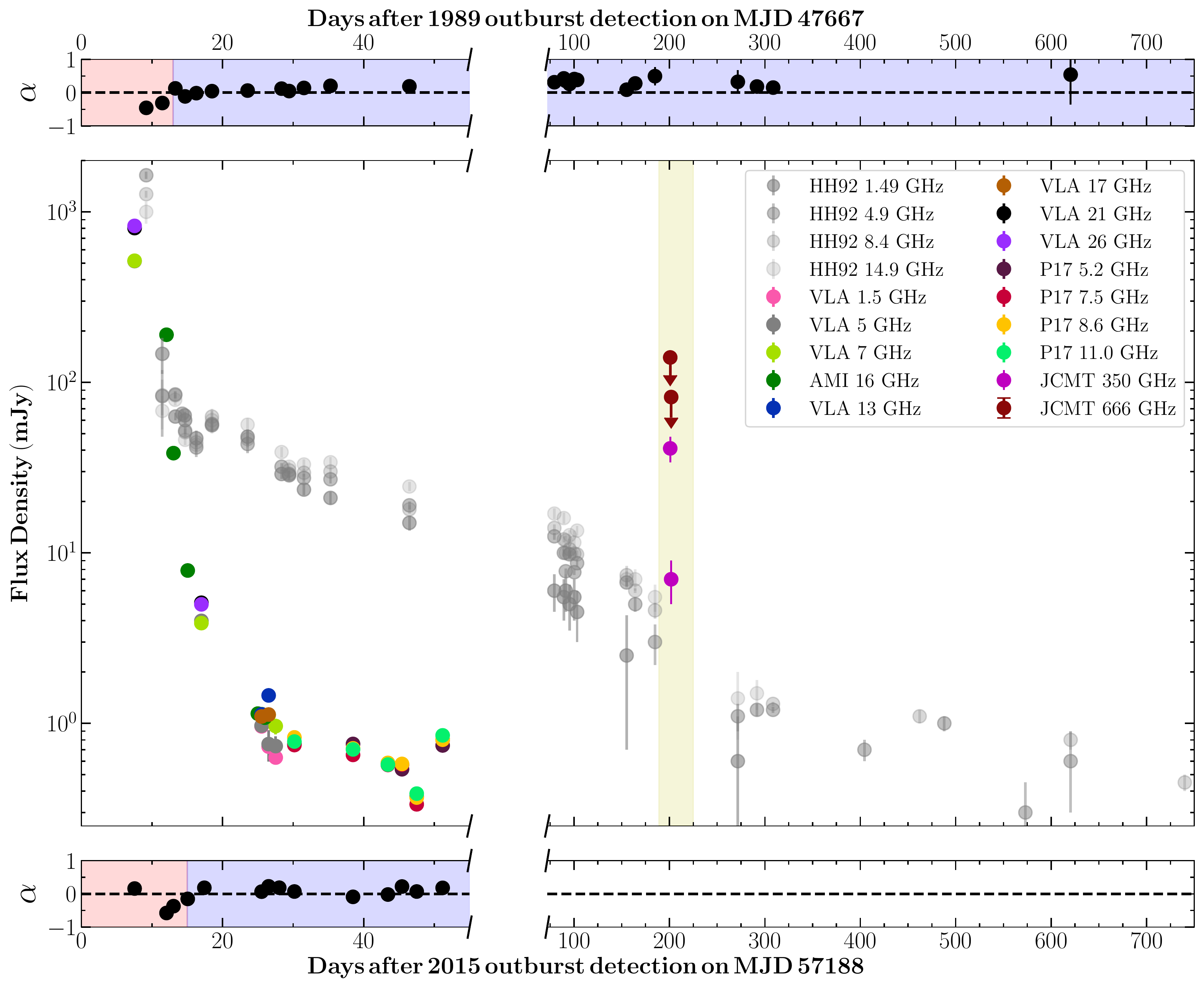}
 \caption{\label{fig:lc1989} Radio frequency light curves of the 1989 (gray-scale data points) and 2015 (coloured data points) outbursts of V404 Cyg. The 1989 radio frequency data are taken from \citealt{hanhj92} (HH92), and we supplement our 2015 radio frequency coverage with the measurements reported in \citealt{plot16} (P17). The {\sl top} and {\sl bottom} panels indicate the radio spectral indices (in epochs where at least two radio bands were observed) in 1989 and 2015, respectively. The shaded regions in the {\sl top} and {\sl bottom} panels represent the time periods in which the radio emission was likely dominated by jet ejecta emission (red; $\alpha<0$) or compact jet emission (blue; $\alpha\geq0$). We also include our JCMT sub-mm coverage of the 2015 mini-outburst in this plot, where the yellow shading in the {\sl middle} panel indicates the duration of the 2015 mini-outburst period (MJD 57377--57413; \citealt{mun16b}). While both outbursts reach similar peak intensities and show similar radio spectral index evolution, the radio emission from the 2015 outburst appears to decay much faster than in the 1989 outburst.
}
\end{figure*}

Measuring a spectral break in the optical bands in the V404 Cyg jet spectrum could also have important implications regarding the jet contribution to the X-ray emission in this system.
In particular, with such a high spectral break frequency, the optically thin synchrotron emission from the jet could be dominating the emission in the X-ray bands (e.g., in XTE J1550--564 the jet has been shown to dominate the X-ray bands during the outburst decay; \citealt{rus10}). In \S\ref{sec:specb} and Figure~\ref{fig:seds} we have shown that a synchrotron spectrum (with a cooling break between the UV and X-ray bands) extrapolated to the X-ray bands can reasonably reproduce the X-ray fluxes observed during the decay of this V404 Cyg outburst.
This indicates that the jet could be producing a large fraction of the X-ray flux at this point in the outburst. However, to confirm this theory, a more detailed X-ray analysis (possibly examining hard/soft lags, reflection features, or the presence of iron lines) would need to be performed to verify that the X-ray emission is indeed more consistent with synchrotron from a jet rather than Comptonization in a hot inner flow. Such an X-ray analysis is beyond the scope of this work. {However, we note that X-ray spectral studies \citep{motta17} of this stage of the outburst find photon indexes ($\Gamma\sim 1.5-1.7$) consistent with our estimated optically thin synchrotron spectral indices ($\alpha\sim-0.6\,\mbox{--}\,-0.8$, where $\Gamma=1-\alpha$). This suggests that the X-ray spectrum may show a similar slope to the optically thin part of the compact jet spectrum, and thus be indicative of a synchrotron origin for this X-ray emission.}

Further, we see limited evidence for evolution in the broad-band spectra across our sampled epochs during the outburst decay. For instance, the optically thin spectral index stays relatively constant (within error) across the epochs where it is measured, the optically thick spectral index may only flatten slightly over time, and the spectral break resides in the near-IR/optical bands across all of our sampled epochs (see Table~\ref{table:sed_tab}). Additional radio observations occurring after our sampled epochs (in late July and early August) also show a similar trend, where the shape of the radio spectrum (i.e., spectral index) remains relatively constant over time \citep{plot16}. Limited jet spectral evolution may suggest that the jet properties are not changing significantly as the jet emission fades during these epochs (e.g., the jet spectral shape can be sensitive to many parameters, such as the magnetic field strength at the base of the jet, jet geometry, inclination of the system, the particle acceleration process, and the electron energy distribution injected into the jets; \citealt{peer09,kai06,jam10,mal14}). {Interestingly, \citealt{chpoo17} have reported the detection of a spectral break in the radio band at $\nu_{\rm break}=1.8$ GHz on MJD 57199/57200. This finding is consistent with our previous work \citep{tetarenkoa17}, where we found evidence of a compact jet with a spectral break between $0.341<\nu_{\rm break}<5.25$ GHz on MJD 57195. Therefore, while we observe little spectral break evolution in broad-band spectra sampling epochs later than MJD 57202 in this work, combining these two results may suggest that the spectral break rapidly shifted from the radio to near-IR bands over the span of a few days (between MJD 57199--57202).} 

Moreover, we find that the compact jet emission during the outburst decay can also be highly variable (on minute to hour timescales), similar to the conditions observed in the quiescent V404 Cyg jet. While the variability amplitude at radio/sub-mm frequencies appears to follow an average trend, where the variability amplitude decreases as the intensity of the jet emission decreases, interestingly we also find that the variability amplitude can sporadically increase in certain time periods.
For example, on MJD\,57204 the sub-mm emission displays a large variability amplitude of $F_{\rm var}\sim85\%$, while the epochs taken $\sim 24$ hours prior to and following the MJD\,57204 epoch, show little to no variability (see Figure~\ref{fig:trlc_jul0102}). \cite{maitra17} have suggested that strong optical frequency variability (also probing the jet base region, but closer to the black hole than probed by sub-mm frequencies) occurring a few days earlier on MJD\,57200, could be caused by a disruption in the feeding of the jet. In this situation, the mass outflow rate changes sporadically in response to a change in the mass inflow rate through the disc. Alternatively, the high variability amplitude detected on MJD\,57204 could be tracing the re-establishment of the compact jet, following the last major flaring episodes occurring a few days earlier.
In either case, if the jet flow was unstable in the MJD\,57204 epoch, this could also explain the origin of the time lags we detect between the radio bands in the MJD\,57205 epoch. In particular, these time lags could be tracking a disturbance in the jet flow, that was injected into the jet base sometime between the MJD\,57204/57205 epochs, and has since propagated downstream (to lower frequencies) in the jet.

\subsection{Comparison to the 1989 Outburst}
V404 Cyg underwent three known outbursts prior to 2015: 1938 (when the optical counterpart was observed and originally identified as Nova Cygni; \citealt{wagn89}), 1956 (as discovered on photographic plates; \citealt{r87}), and 1989 (when the transient X-ray counterpart was first identified by the Ginga satellite; \citealt{ma89}). The 1989 outburst was monitored at multiple frequencies, including X-ray, optical, and radio frequencies \citep{hanhj92}. While both the 1989 and 2015 outburst have been shown to exhibit similar X-ray behaviour (e.g., 
bright X-ray flaring activity that is not always intrinsic to the source), a detailed comparison between the radio jet behaviour during these different outbursts has not yet been presented.
Radio frequencies offer the distinct advantage of providing a cleaner view of the system, when compared to the X-ray regime, as the high column density/external obscuration effects \citep{motta17b} do not apply in the radio regime.
As such, given our well-sampled radio coverage, in this section we compare the radio jet behaviour between the 1989 and 2015 outbursts (see Figure~\ref{fig:lc1989} for the radio frequency light curves and radio spectral indices for both outbursts).

The radio jet behaviour in the 1989 and 2015 outbursts display many similarities; both outbursts reach similar peak intensities ($\sim1$ Jy), the spectral indices show similar evolution (progresses from steep, to flat, to inverted, in $\sim20$ days), both outbursts spend
$\sim 15$ days in a high luminosity flaring accretion state before transitioning into a hard accretion state (where the jet emission is dominated by a compact jet; red and blue shading in Figure~\ref{fig:lc1989}), the radio emission remained variable throughout the outburst (regardless of flux level), and there are hints of coupling between radio, optical, and X-ray emission \citep{hanhj92,tetarenkoa17,plot16}. However, the radio emission in the 2015 outburst decays significantly faster than the radio emission in the 1989 outburst ($\sim30$ days in 2015 vs. $\sim300$ days in 1989, to reach sub-mJy levels), and no mini-outburst is observed at radio frequencies in 1989 \citep{mun16b}. However, given the duration of the 2015 mini-outburst, and the sampling timescale between the 1989 radio epochs, it is entirely possible that such a mini-outburst was simply missed in 1989. Interestingly, the 2015 outburst also decayed much more quickly in X-rays when compared to the 1989 outburst ($\sim60$ days in 2015 vs $\sim160$ days in 1989; \citealt{ter94,oos97,zy99,plot16}).

Several works \citep{galfenpol03,corb08,gallo14,plot16,gallo18} have shown that the radio luminosity of V404 Cyg is linked to the X-ray luminosity through a robust disc-jet coupling relationship (radio luminosity is proportional to the X-ray luminosity; $L_R\propto L_X^\beta$, where $\beta\sim0.54$ for V404 Cyg). 
This relationship was shown to hold across a wide range of X-ray luminosities ($L_X\sim 10^{32}-10^{37}\,{\rm erg\,s}^{-1}$), and is valid for both the 1989 and 2015 outbursts \citep{plot16}. Therefore, as the X-ray luminosity can be thought of a proxy for mass accretion rate, the more rapid drop in radio luminosity in the 2015 outburst could be the result of a more rapid drop in the average mass accretion rate after the peak of the outburst, when compared to the 1989 outburst.
\cite{mun16} have suggested that the strong accretion disc wind detected during the 2015 outburst may be a factor that regulates outburst duration, as these winds can significantly deplete the mass in the accretion disc (and potentially cause drastic and rapid changes in mass accretion rate). As such, it is possible that the faster decay seen in our radio light curves in the 2015 outburst indicates that the mass loss rate in the winds was much higher in the 2015 outburst, leaving less matter in the disc to fuel the jets. Along the same lines, the length of the quiescent period prior to each outburst may also be a contributing factor to the more rapid radio decay. In particular, as the quiescent period prior to the 1989 outburst was longer than the quiescent period prior to the 2015 outburst (33 vs 26 years), the system had more time to build up mass in the disc (to fuel the jets) before the 1989 outburst.

\subsection{The December 2015 Mini-outburst}

Following the June 2015 outburst of V404 Cyg, renewed X-ray activity was detected from the system in December 2015 (MJD 57377, $\sim189$ days after the first detection of the June outburst; \citealt{lip15,trus15,beard15,mal15,mott16a}). Recent work \citep{mun16b,kaj18} has shown that while this ``mini-outburst" was in general fainter across all sampled frequencies when compared to the June outburst, it showed similar phenomenology; strong flaring activity, fast accretion disc wind, highly variable column density.

This mini-outburst phenomenon has been seen in other BHXB sources following bright outbursts (e.g.,  GRO J0422+32, XTE J1650-500, MAXI J1659-152, GRS 1739-278; \citealt{yan17} and references within). However, with the exception of one source (Swift J1753.5-0127; \citealt{plotk17}), these mini-outbursts have only been monitored at X-ray and optical frequencies.

\begin{figure}
\begin{center}
 \includegraphics[width=1\columnwidth]{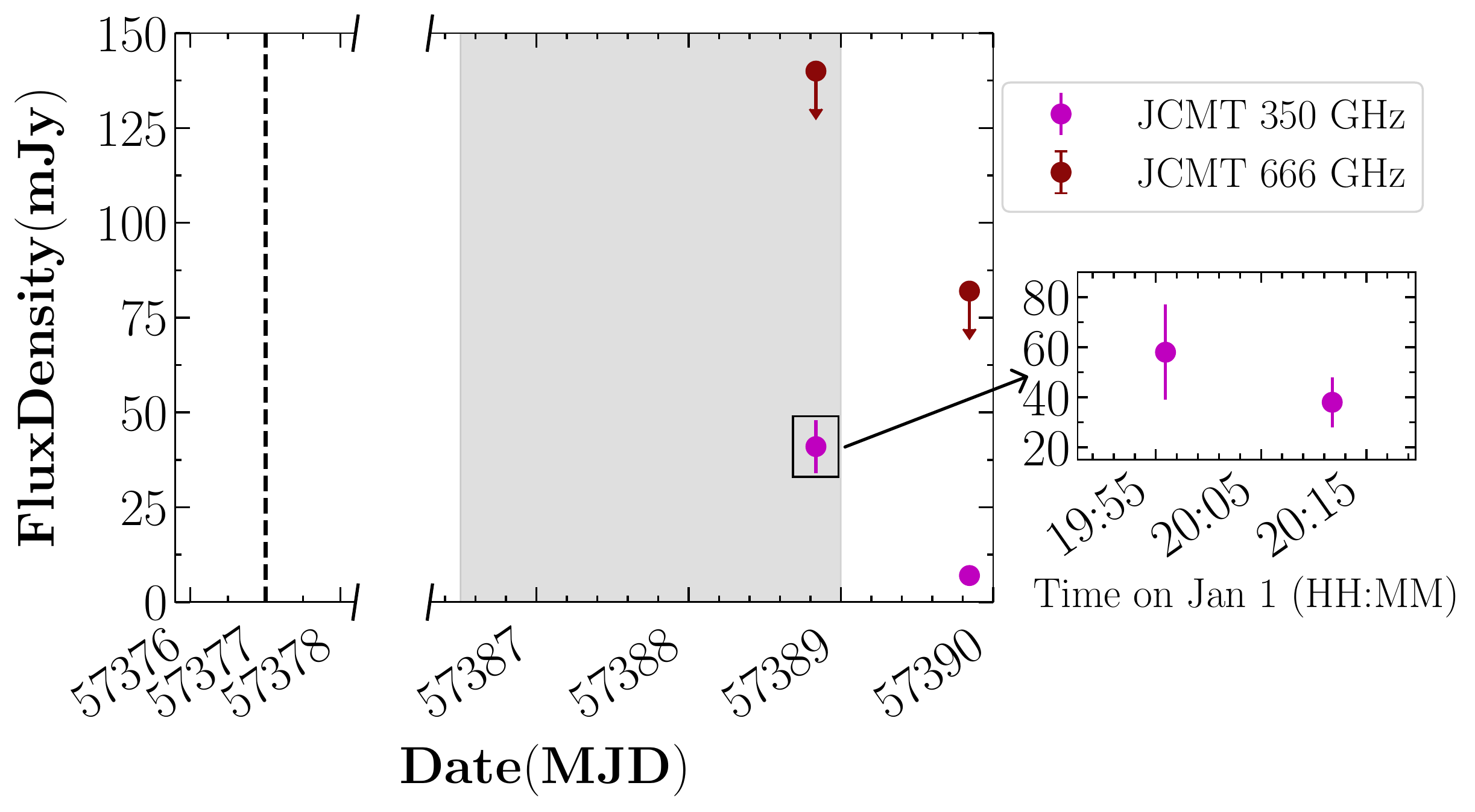}
 \caption{\label{fig:reburst} JCMT sub-mm light curves of V404 Cyg during the December 2015 mini-outburst. The vertical dotted line indicates the start of the mini-outburst on MJD 57377, while the shaded grey region indicates the most active flaring phase of the mini-outburst (MJD 57386.5--57389; as identified by \citealt{mun16b}). The {\sl right panel} displays a time-resolved analysis of the 2016 January 1 (MJD 57388) observation, where we split the 30 minute scan into two segments. The decreasing flux trend observed in the daily light curve ({\sl main panel}) and time resolved measurements ({\sl right panel}), as well as the timing of our measurements near the end of the active flaring period, suggests that we may have caught the tail end of a brighter flare in these JCMT sub-mm observations.
}
\end{center}
\end{figure}

\cite{mun16b} present AMI radio monitoring of this mini-outburst period, where they detect flaring radio emission for a 10 day period following the mini-outburst detection. This flaring period ended with a large radio flaring episode on December 31/January 1, after which the radio emission began to decay (similar to the final radio flare on June 26 in the main outburst). We obtained JCMT SCUBA-2 sub-mm observations on January 1 and 2, in which we observe a decreasing flux density trend in both the time-resolved measurements on January 1 and between the two JCMT epochs (see Figure~\ref{fig:reburst}). This trend, combined with the timing of our measurements near the large radio flare, suggests that we may have caught the tail end of a brighter flare in our JCMT observations. Radio and sub-mm flaring accompanied repeated jet ejection events during the main June outburst (\citealt{tetarenkoa17} and Figure~\ref{fig:trlc_jun2630}). Therefore, our measurements corroborate the theory suggested by \cite{mun16b}, that despite the fainter nature of the mini-outburst, V404 Cyg was still launching discrete jet ejecta during this time period, and in turn, jet ejecta are not exclusive to the highest luminosity states in V404 Cyg.
 
\section{Summary}
\label{sec:sum}
In this paper, we present the results of our multi-frequency monitoring observations of the 2015 outburst of the BHXB V404 Cyg. We observed the source at radio and mm/sub-mm frequencies, with the VLA, AMI, SMA, JCMT, and NOEMA, and collected publicly available OIR, UV, and X-ray measurements to supplement our coverage.
With this well-sampled data set, we created detailed diagnostics of the jet emission in V404 Cyg, tracking the spectral and variability properties throughout different stages of the outburst (e.g., time-resolved light curves and spectral indices, broad-band spectra, CCFs, variability amplitude measurements).

Using these diagnostics we find that the jet emission was originally dominated by emission from discrete jet ejecta during the brightest stages of the outburst. These ejecta appeared to become fainter, slower, less frequent, and less energetic, before the emission abruptly (over 1--2 days) transitioned to being dominated by a compact jet.
While the broad-band spectrum of this compact jet showed very little evolution throughout the outburst decay (despite the intensity of the jet emission fading by an order of magnitude), the emission still remained intermittently variable at mm/sub-mm frequencies. Further, through phenomenological modelling of the broad-band emission from this compact jet, we directly detect the optically thick to thin synchrotron jet spectral break in the near-IR and optical bands ($\sim2-5\times10^{14}$ Hz), and postulate that the compact jet could have been significantly contributing to the X-ray emission observed during the outburst decay.

Additionally, we compared the radio jet emission throughout the 2015 and previous 1989 outbursts. While the radio jet emission in both outbursts show many similarities (e.g., peak flux, spectral index evolution), we show that the radio emission in the 2015 outburst decayed significantly (${\sim10}$ {times}) faster than in the 1989 outburst. We postulate that, given the robust disc-jet coupling relationship found between these two outbursts \citep{plot16}, this difference could indicate that the average mass accretion rate dropped (possibly due to the strong accretion disc wind) much quicker following the peak of the 2015 outburst, when compared to the 1989 outburst.

Lastly, we report on sub-mm observations during the December 2015 mini-outburst of V404 Cyg. These sub-mm observations display a decreasing flux trend, that most likely samples the tail end of a larger flaring episode. As sub-mm flaring coincided with jet ejection episodes during the main outburst, these observations support previous claims \citep{mun16b} that, similar to the main outburst, the source was most likely launching powerful jet ejecta during this period of renewed activity.

Overall, our work here demonstrates the importance of simultaneous, multi-frequency, time-resolved observations, to fully understand the rapidly evolving jet sources in BHXBs. 

\section*{Acknowledgements}
The authors thank the anonymous referee for constructive feedback that improved the manuscript.
We wish to sincerely thank all the National Radio Astronomy Observatory, SMA, JCMT, and IRAM NOEMA staff that supported us during this observing campaign.
The authors also offer a special thanks to the IRAM NOEMA Director, Karl-Fredrich Schuster, for granting our NOEMA DDT request.
AJT thanks Craig Heinke and Bailey Tetarenko for useful discussions on accretion disc contributions in broad-band spectra of X-ray binaries.  AJT is supported by an Natural Sciences and Engineering Research Council of Canada (NSERC) Post-Graduate Doctoral Scholarship (PGSD2-490318-2016). AJT and GRS are supported by NSERC Discovery Grants. JCAMJ is the recipient of an Australian Research Council Future Fellowship (FT140101082). SBM acknowledges support from VICI grant Nr. 639.043.513/520, funded by the Netherlands Organisation for Scientific Research (NWO). TDR acknowledges support from the NWO Veni Fellowship, grant number 639.041.646. DA acknowledges support from the Royal Society. This work is based on observations carried out under project numbers S15DE \& D15AB with the IRAM NOEMA Interferometer. IRAM is supported by INSU/CNRS (France), MPG (Germany) and IGN (Spain). The National Radio Astronomy Observatory is a facility of the National Science Foundation operated under cooperative agreement by Associated Universities, Inc. We thank the engineers and staff of the 
Mullard Radio Astronomy Observatory for maintenance and operation 
of AMI, which is supported by the Universities of Cambridge and 
Oxford. The AMI telescope acknowledges support from the European 
Research Council under grant ERC-2012-StG-307215 LODESTONE. The Sub-millimeter Array is a joint project between the Smithsonian Astrophysical Observatory and the Academia Sinica Institute of Astronomy and Astrophysics, and is funded by the Smithsonian Institution and the Academia Sinica. 
The James Clerk Maxwell Telescope is operated by the East Asian Observatory on behalf of The National Astronomical Observatory of Japan; Academia Sinica Institute of Astronomy and Astrophysics; the Korea Astronomy and Space Science Institute; the Operation, Maintenance and Upgrading Fund for Astronomical Telescopes and Facility Instruments, budgeted from the Ministry of Finance (MOF) of China and administrated by the Chinese Academy of Sciences (CAS), as well as the National Key R\&D Program of China (No. 2017YFA0402700). Additional funding support is provided by the Science and Technology Facilities Council of the United Kingdom and participating universities in the United Kingdom and Canada.
The authors also wish to recognize and acknowledge the very significant cultural role and reverence that the summit of Mauna Kea has always had within the indigenous Hawaiian community.  We are most fortunate to have the opportunity to conduct observations from this mountain.
We acknowledge with thanks the variable star observations from the AAVSO International Database contributed by observers worldwide and used in this research.



\bibliography{ABrefList} 



\appendix
\section{Observation Setup}
In this section we provide details on the correlator and array setup of all of our radio through mm/sub-mm interferometric observations; VLA (Table~\ref{table:vla_obs}), NOEMA (Table~\ref{table:noema_obs}), SMA (Table~\ref{table:sma_obs}).

\section{Observational Data}
In this section we provide data tables of all the day-timescale multi-frequency photometry measurements presented in this work; radio (Table~\ref{table:vla_flux}), mm/sub-mm (Table ~\ref{table:submm_flux}), and OIR/UV/X-ray (Table~\ref{table:other_flux}).

\section{Jet Modelling Results}
\label{sec:vdl_extra}
\begin{figure*}
   \centering
   \includegraphics[height=1.5\columnwidth]{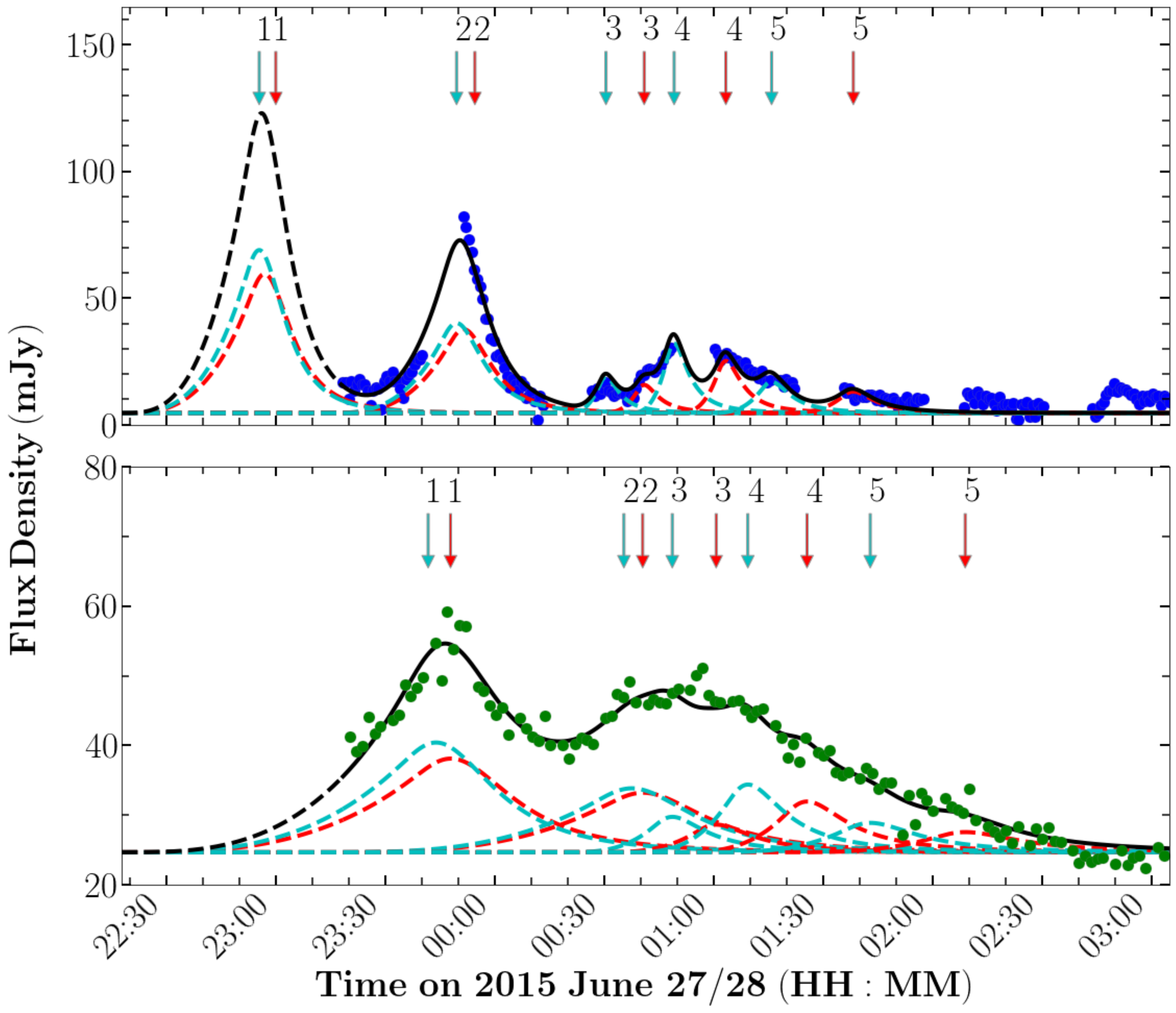} \\
 \caption{\label{fig:vdl_update}  Radio (AMI 16 GHz; {\sl bottom panel}) and mm/sub-mm (NOEMA 140 GHz; {\sl top panel}) light curves of V404 Cyg on 2015 
 June 27/28 (MJD 57200/57201). In both panels, the black lines represents our predicted best fit jet model at each frequency, and the dotted lines indicate the approaching (cyan) and receding (red) components of the individual ejection events. The arrows at the top of each panel (cyan for approaching, red for receding) identify which flares correspond to which ejection number from Table~\ref{table:vdl_table}.
 }
 \end{figure*}

\begin{figure*}
   \centering
   \includegraphics[height=1.5\columnwidth]{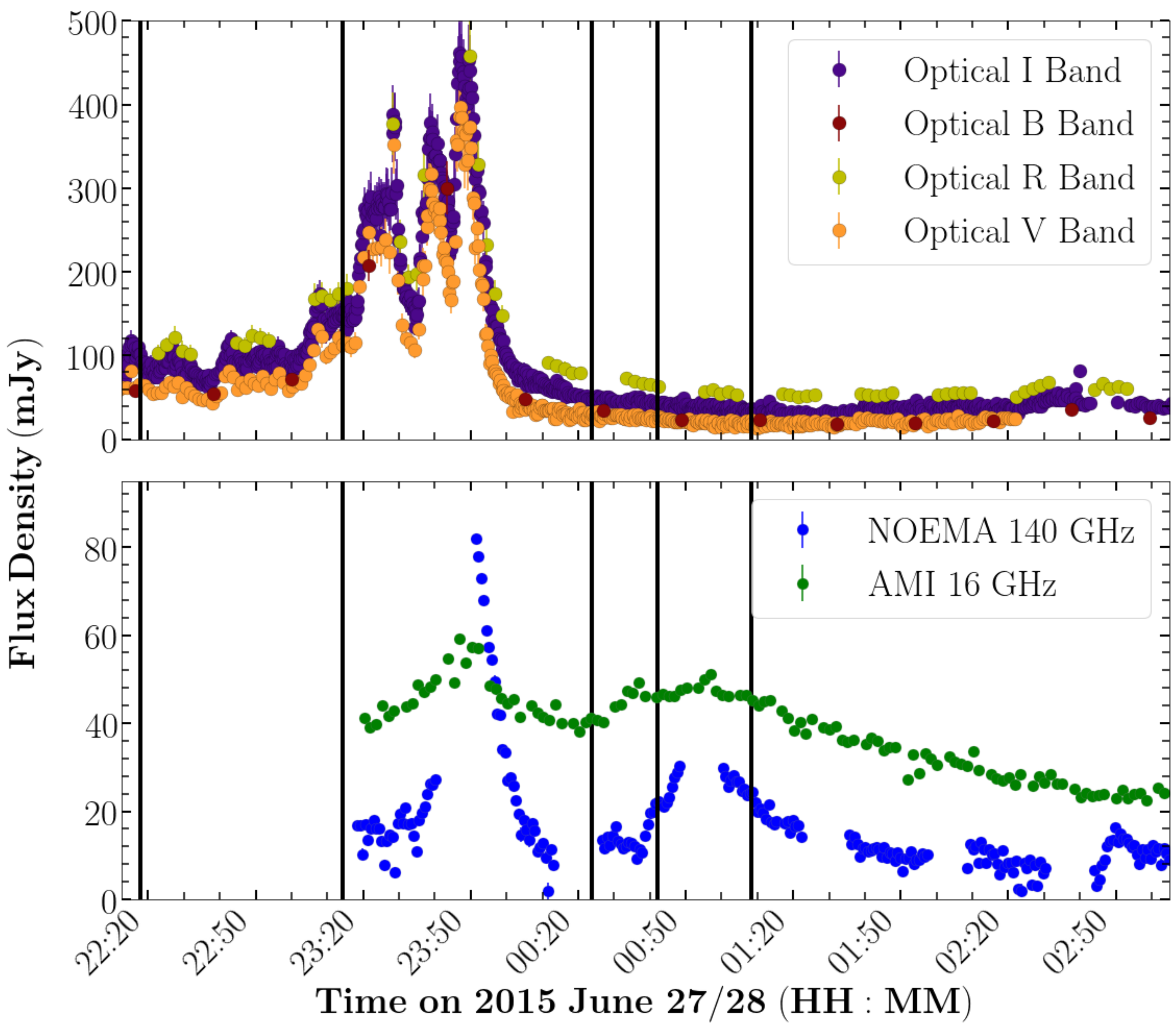} \\
 \caption{\label{fig:vdl_ejtimes} {Zoomed in version of Figure~\ref{fig:trlc_jun2630}, showing the optical (I, B. R. and V bands; {\sl top panel}), as well as radio (AMI 16 GHz) and mm/sub-mm (NOEMA 140 GHz; {\sl bottom panel}) light curves of V404 Cyg on 2015 
 June 27/28 (MJD 57200/57201). The vertical black lines indicate the predicted ejection times for each jet ejection event we model (see Table~\ref{table:vdl_table}). While ejection 2 coincides with the beginning of a large optical flaring complex, the ejection times for later events do not seem to have optical counterparts (similar to what was seen during our previous modelling of flaring on MJD 57195; \citealt{tetarenkoa17})}.
 }
 \end{figure*}

{In this section we show additional figures pertaining to our light curve modelling. Figure~\ref{fig:vdl_update} displays an alternate version of Figure~\ref{fig:vdl}, in which we decompose the total jet model into the individual approaching and receding components. Figure~\ref{fig:vdl_ejtimes} displays a zoomed in version of Figure~\ref{fig:trlc_jun2630}, where we also indicate the ejection times predicted from our modelling.}

\renewcommand\tabcolsep{15pt}
 \begin{table*}
\caption{VLA Observations Summary}\quad
\centering
\begin{tabular}{ cccccc }
 \hline\hline
  {Date} &{MJD}&{Sub-array}&{Scans}&{Band(s)$^a$}&{Number of}\\
  {(2015)}&&{}&{(UTC)}&&{Antennas}\\[0.15cm]
  \hline

July 02&57205&A&10:02:14--14:01:32&C-K-C&14\\[0.15cm]
&&B&10:05:14--14:01:32&K-C-K&13\\[0.15cm]
July 10 & 57213&A&12:03:02--13:01:50&C&8\\[0.15cm]
&&B&12:04:02--13:01:52&L&9\\[0.15cm]
&&C&12:02:02--13:01:50&Ku&9\\[0.15cm]
July 11 & 57214&A&13:29:52--14:27:40&L&9\\[0.15cm]
&&B&13:28:52--14:27:40&C&8\\[0.15cm]
&&C&13:27:52--14:27:40&Ku&9\\[0.15cm]
July 12 & 57215&A&12:04:02--13:01:52&L&8\\[0.15cm]
&&B&12:03:02--13:01:50&C&9\\[0.15cm]
&&\phantom{0}C$^{b}$&\dots&Ku&8\\[0.15cm]\hline
\end{tabular}\\
\begin{flushleft}
{$^a$ This column indicates the frequency bands observed with each sub-array. When multiple bands are present, the entry in this column indicates the temporal sequence with which the bands were observed by the sub-array.}\\
{$^{b}$ A technical error occurred during observing and no data was taken for this sub-array.}\\
\end{flushleft}
\label{table:vla_obs}
\end{table*}
\renewcommand\tabcolsep{6pt}

\renewcommand\tabcolsep{15pt}
 \begin{table*}
\caption{NOEMA Observations Summary}\quad
\centering
\begin{tabular}{ cccccccc }
 \hline\hline
  {Date} &{MJD}&{Array Config.}&{Scans}&{Band }&{Number of}\\
  {(2015)}&&&{(UTC)}&&{Antennas}\\[0.15cm]
  \hline
Jun 26&57199&6Dq-E03+E12&22:24:02--23:12:58&3 mm&6\\[0.15cm]
Jun 27 &57200&6Dq-E03+E12&00:15:35--02:42:31&2 mm&6\\[0.15cm]
Jun 27/28 &57200/57201&6Dq-E03+E12&23:17:37--03:05:34&2 mm&6\\[0.15cm]
Jun 29&57202&6Dq&22:13:50--23:02:47&3 mm&6\\[0.15cm]
Jun 30 &57203&6Dq&00:14:51--02:55:10&2 mm&6\\[0.15cm]
Jul 12&57215&6Dq&22:03:27--22:52:23&3 mm&6\\[0.15cm]
Jul 12/13 &57215/57216&6Dq&23:45:19--02:12:09&2 mm&6\\[0.15cm]
\hline
\end{tabular}\\
\label{table:noema_obs}
\end{table*}
\renewcommand\tabcolsep{6pt}

\renewcommand\tabcolsep{3pt}
 \begin{table*}
\caption{SMA Observations Summary$^a$}\quad
\centering
\begin{tabular}{ cccccccc }
 \hline\hline
  {Date} &{MJD}&{Array Config.}&{Scans}&{Total  Bandwidth}&{Correlator}&{Correlator Setup}&{Number of}\\
  {(2015)}&&&{(UTC)}&{(GHz)}&& ({$N_{\rm spw}$, $N_{\rm C}$, $\Delta\, \nu$ MHz)$^b$}&{Antennas}\\[0.15cm]
  \hline
Jun 16&57189&sub-compact&17:22:33--18:40:43&8.32&ASIC+SWARM&\phantom{0}(48, 128, 0.8125) + (2, 128, 13)$^c$&7\\[0.15cm]
Jun 17&57190&sub-compact&13:25:04--17:03:14&4.992&ASIC&(48, 128, 0.8125)&6\\[0.15cm]
July 01&57204&compact&07:50:28--17:18:53&2.080&ASIC&(20, 32, 3.250)&6\\[0.15cm]
July 02&57205&compact&09:26:57--14:30:42&2.496&ASIC&\phantom{0}(24, 32, 3.250)$^d$&6\\[0.15cm]\hline
\end{tabular}\\
\begin{flushleft}
{$^a$ Our SMA project is a ToO program, with highly constrained start times, needed to obtain simultaneous observations with other facilities. Therefore, due to the continuum nature of our observations, our program is often run immediately before or after other SMA observing programs, which results in the wide variety of correlator setups seen here.}\\
{$^b$ The correlator setup; number of spectral windows, number of channels, and channel width, for each of the two side-bands.}\\
{$^c$ The SWARM correlator had a fixed resolution of 101.6 kHz per channel, with 16383 channels for each SWARM spectral window.  Given the continuum nature of these observations, we performed spectral averaging, to yield 128 13 MHz channels in both SWARM spectral windows (matching the number of channels in the ASIC spectral windows), to make it easier to combine ASIC and SWARM data.}\\
{$^d$ Three of the 24 spectral windows were set up with a higher spectral resolution in this observation; 512 0.203 MHz channels.  As such, we spectrally averaged these channels to match the other spectral windows with 32 3.250 MHz channels.}\\
\end{flushleft}
\label{table:sma_obs}
\end{table*}
\renewcommand\tabcolsep{6pt}

\renewcommand\tabcolsep{2pt}
 \begin{table}
\caption{Flux Densities of V404 Cyg at Radio Frequencies}\quad
\centering
\begin{tabular}{ ccccc }
 \hline\hline
 {Telescope}& {Date} &{MJD}&{Freq.}&{Flux Density$^a$}\\
  &{(2015)}&&{(GHz)}&{(mJy)}\\[0.15cm] \hline
VLA&June 22&57195&5.25& $514.5\pm5.1$\\[0.15cm]
VLA&June 22&57195&7.45& $516.9\pm5.2$\\[0.15cm]
VLA&June 22&57195&20.8& $803.5\pm18.1$\\[0.15cm]
VLA&June 22&57195&25.9& $827.7\pm18.3$\\[0.15cm]
AMI&June 26/27& 57199/57200&16.0&$190.23\pm0.03$\\[0.15cm]
AMI&June 27/28& 57200/57201&16.0&$38.44\pm0.03$\\[0.15cm]
AMI&June 29/30& 57202/57203&16.0&$7.88\pm0.02$\\[0.15cm]
GMRT$^\dagger$&July 01&57204&1.280& $6.39\pm0.67$\\[0.15cm]
GMRT$^\dagger$&July 01&57204&0.610& $8.88\pm0.94$\\[0.15cm]
GMRT$^\dagger$&July 01&57204&0.235& $13.4\pm2.4$\\[0.15cm]
VLA&July 02&57205&5.25& $3.99\pm0.06$\\[0.15cm]
VLA&July 02&57205&7.45& $3.87\pm0.05$\\[0.15cm]
VLA&July 02&57205&20.8& $5.10\pm0.16$\\[0.15cm]
VLA&July 02&57205&25.9& $4.99\pm0.15$\\[0.15cm]
VLA&July 10&57213&1.52& $0.96\pm0.09$\\[0.15cm]
VLA&July 10&57213&5.24& $0.97\pm0.09$\\[0.15cm]
VLA&July 10&57213&7.45& $1.13\pm0.08$\\[0.15cm]
VLA&July 10&57213&13.5& $1.13\pm0.05$\\[0.15cm]
VLA&July 10&57213&17.4& $1.10\pm0.05$\\[0.15cm]
AMI&July 10& 57213&16.0&$1.14\pm0.07$\\[0.15cm]
GMRT$^\dagger$&July 11&57214&1.280& $0.52\pm0.14$\\[0.15cm]
GMRT$^\dagger$&July 11&57214&0.610& $<0.6$\\[0.15cm]
GMRT$^\dagger$&July 11&57214&0.235& $0.66\pm0.24$\\[0.15cm]
VLA&July 11&57214&1.52& $0.73\pm0.05$\\[0.15cm]
VLA&July 11&57214&5.24& $0.76\pm0.16$\\[0.15cm]
VLA&July 11&57214&7.45& $1.12\pm0.10$\\[0.15cm]
VLA&July 11&57214&13.5& $1.46\pm0.07$\\[0.15cm]
VLA&July 11&57214&17.4& $1.13\pm0.07$\\[0.15cm]
AMI&July 11& 57214&16.0&$1.08\pm0.07$\\[0.15cm]
GMRT$^\dagger$&July 12&57215&1.280& $<4.7$\\[0.15cm]
GMRT$^\dagger$&July 12&57215&0.610& $0.52\pm0.12$\\[0.15cm]
GMRT$^\dagger$&July 12&57215&0.235& $0.50\pm0.19$\\[0.15cm]
VLA&July 12&57215&1.52& $0.63\pm0.04$\\[0.15cm]
VLA&July 12&57215&5.24& $0.74\pm0.11$\\[0.15cm]
VLA&July 12&57215&7.45& $0.96\pm0.10$\\[0.15cm]
VLA&July 12$^\star$&57215&13.5& \dots\\[0.15cm]
VLA&July 12$^\star$&57215&17.4& \dots\\[0.15cm]
\hline
\end{tabular}\\
\begin{flushleft}
{$^a$ The VLA measurements include the standard VLA systematic errors.}\\
{$^\star$ A technical error occurred during observing, and thus no data was taken for this sub-array.}\\
{$^\dagger$ Giant Metrewave Radio Telescope (GMRT) data are taken from \citet{chpoo17}.}\\
\end{flushleft}
\label{table:vla_flux}
\end{table}
\renewcommand\tabcolsep{6pt}

\renewcommand\tabcolsep{2pt}
 \begin{table}
\caption{Flux Densities of V404 Cyg at mm/sub-mm Frequencies}\quad
\centering
\begin{tabular}{ ccccc }
 \hline\hline
 {Telescope}& {Date} &{MJD}&{Freq.}&{Flux Density}\\
  &{(2015)}&&{(GHz)}&{(mJy)}\\[0.15cm] \hline
SMA&June 16&57189&220.25& $102.0\pm1.6$\\[0.15cm]
SMA&June 16&57189&230.25& $72.9\pm1.5$\\[0.15cm]
SMA&June 17 &57190&220.25& $17.5\pm0.6$\\[0.15cm]
SMA&June 17&57190&230.25& $15.7\pm0.6$\\[0.15cm]
JCMT&June 17&57190&350& $7.9\pm2.1$\\[0.15cm]
JCMT&June 17&57190&\phantom{0}666$^\star$& $<122$\\[0.15cm]
SMA&June 22&57195&220.25& $878.0\pm32.0$\\[0.15cm]
SMA&June 22&57195&230.25& $872.0\pm32.0$\\[0.15cm]
JCMT&June 22&57195&350& $932.8\pm6.9$\\[0.15cm]
JCMT&June 22&57195&666& $988.6\pm30.0$\\[0.15cm]
NOEMA&June 26/27&57199/57200&97.5&$70.16\pm0.09$\\[0.15cm]
NOEMA&June 26/27&57199/57200&140&$48.62\pm0.07$\\[0.15cm]
NOEMA&June 27/28&57200/57201&140&$16.80\pm0.05$\\[0.15cm]
NOEMA&June 29/30&57202/57203&97.5&$5.32\pm0.09$\\[0.15cm]
NOEMA&June 29/30&57202/57203&140&$5.87\pm0.07$\\[0.15cm]
SMA&July 01 &57204&220.25& $5.2\pm1.2$\\[0.15cm]
SMA&July 01&57204&230.25& $5.8\pm1.2$\\[0.15cm]
SMA&July 02 &57205&220.25& $8.9\pm2.2$\\[0.15cm]
SMA&July 02&57205&230.25& $5.5\pm2.5$\\[0.15cm]
JCMT&July 02 &57205&350& $10.0\pm1.4$\\[0.15cm]
JCMT&July 02 &57205&\phantom{0}666$^\star$& $<80$\\[0.15cm]
NOEMA&July 12/13&57215/57216&97.5&$1.38\pm0.09$\\[0.15cm]
NOEMA&July 12/13&57215/57216&140&$1.44\pm0.12$\\[0.15cm]
JCMT&\phantom{0}January 01$\dagger$&57388&350&$41\pm7$\\[0.15cm]
JCMT&\phantom{0}January 01$\dagger$&57388&\phantom{0}666$^{\star}$&$<140$\\[0.15cm]
JCMT&\phantom{0}January 02$\dagger$&57389&350&$7\pm2$\\[0.15cm]
JCMT&\phantom{0}January 02$\dagger$&57389&\phantom{0}666$^{\star}$&$<82$\\[0.15cm]
\hline
\end{tabular}\\
\begin{flushleft}
{$^\star$ Note that data in the 666 GHz (450$\mu m$) band was obtained simultaneously with the 350 GHz (850 $\mu m$) band, but the source was not always significantly detected at 666 GHz. The values reported here represent $3\sigma$ upper limits.}\\
{$\dagger$ These data were taken during the late December 2015/January 2016 mini-outburst of V404 Cyg.}\\
\end{flushleft}
\label{table:submm_flux}
\end{table}
\renewcommand\tabcolsep{6pt}

\renewcommand\tabcolsep{8pt}
 \begin{table*}
\caption{Flux Densities of V404 Cyg at IR/Optical/UV/X-ray Frequencies}\quad
\centering
\begin{tabular}{ cccccc }
 \hline\hline
 {Band$^a$}& {Date} &{MJD}&{Freq.}&{Flux Density}&{Ref.$^b$}\\
  &{(2015)}&&{(GHz)}&{(mJy)}\\[0.15cm] \hline
R&June 16&57189&4.56$\times 10^5$&$212.2\pm22.9$& [5]\\[0.05cm]
J&June 26/27&57199/57200&2.45$\times 10^5$& $46.6\pm4.9$& [1]\\[0.05cm]
H&June 26/27&57199/57200&1.84$\times 10^5$& $50.5\pm5.0$& [1]\\[0.05cm]
K&June 26/27&57199/57200&1.37$\times 10^5$& $37.6\pm2.9$& [1]\\[0.05cm]
U&June 26/27&57199/57200&8.65$\times 10^5$& $25.7\pm1.0$& [2]\\[0.05cm]
J&June 27/28&57200/57201&2.45$\times 10^5$& $98.7\pm8.6$& [3]\\[0.05cm]
H&June 27/28&57200/57201&1.84$\times 10^5$& $77.9\pm8.5$& [3]\\[0.05cm]
K&June 27/28&57200/57201&1.37$\times 10^5$& $67.4\pm6.6$& [3]\\[0.05cm]
V&June 27/28&57200/57201&5.48$\times 10^5$& $19.1\pm1.9$& [2]\\[0.05cm]
B&June 27/28&57200/57201&6.82$\times 10^5$& $16.1\pm2.4$& [2]\\[0.05cm]
UV1&June 27/28&57200/57201&1.15$\times 10^6$& $2.2\pm9.1$& [2]\\[0.05cm]
J&June 29/30&57202/57203&2.45$\times 10^5$& $30.7\pm0.6$& [4]\\[0.05cm]
H&June 29/30&57202/57203&1.84$\times 10^5$& $25.8\pm0.6$& [4]\\[0.05cm]
K&June 29/30&57202/57203&1.37$\times 10^5$& $22.3\pm0.5$& [4]\\[0.05cm]
I&June 29/30&57202/57203&3.72$\times 10^5$&$30.9\pm2.7$& [5]\\[0.05cm]
V&June 29/30&57202/57203&5.48$\times 10^5$& $19.9\pm2.0$& [2]\\[0.05cm]
B&June 29/30&57202/57203&6.82$\times 10^5$& $18.4\pm2.4$& [2]\\[0.05cm]
U&June 29/30&57202/57203&8.65$\times 10^5$& $11.4\pm2.3$& [2]\\[0.05cm]
UV1&June 29/30&57202/57203&1.15$\times 10^6$& $12.7\pm10.7$& [2]\\[0.05cm]
UV2&June 29/30&57202/57203&1.55$\times 10^6$& $2.9\pm35.2$& [2]\\[0.05cm]
XRT&June 30&57203&1.21$\times 10^9$& $(6.6^{+0.5}_{-0.4})\times10^{-3}$& [6]\\[0.05cm]
V&July 1&57204&5.48$\times 10^5$& $29.8\pm1.7$& [2]\\[0.05cm]
B&July 1&57204&6.82$\times 10^5$& $20.4\pm1.9$& [2]\\[0.05cm]
U&July 1&57204&8.65$\times 10^5$& $13.8\pm2.3$& [2]\\[0.05cm]
UV1&July 1&57204&1.15$\times 10^6$& $5.7\pm7.9$& [2]\\[0.05cm]
UV2&July 1&57204&1.55$\times 10^6$& $41.0\pm33.3$& [2]\\[0.05cm]
J&July 2&57205&2.45$\times 10^5$& $29.9\pm2.6$& [1]\\[0.05cm]
H&July 2&57205&1.84$\times 10^5$& $31.8\pm2.9$& [1]\\[0.05cm]
K&July 2&57205&1.37$\times 10^5$& $24.3\pm2.1$& [1]\\[0.05cm]
I&July 2&57205&3.72$\times 10^5$&$42.5\pm4.0$& [5]\\[0.05cm]
V&July 2&57205&5.48$\times 10^5$& $19.6\pm2.1$& [2]\\[0.05cm]
B&July 2&57205&6.82$\times 10^5$& $18.9\pm2.4$& [2]\\[0.05cm]
U&July 2&57205&8.65$\times 10^5$& $14.5\pm3.1$& [2]\\[0.05cm]
UV1&July 2&57205&1.15$\times 10^6$& $7.3\pm9.7$& [2]\\[0.05cm]
XRT&July 2&57205&1.21$\times 10^9$& $(4.7^{+0.3}_{-0.3})\times10^{-3}$& [6]\\[0.05cm]
I&July 10&57213&3.72$\times 10^5$& $22.3\pm2.0$& [8]\\[0.05cm]
R&July 10&57213&4.56$\times 10^5$& $23.4\pm0.4$& [8]\\[0.05cm]
V&July 10&57213&5.48$\times 10^5$& $14.8\pm0.8$& [8]\\[0.05cm]
U&July 10&57213&8.65$\times 10^5$& $5.1\pm0.6$& [2]\\[0.05cm]
UV1&July 10&57213&1.15$\times 10^6$& $6.1\pm1.3$& [2]\\[0.05cm]
XRT&July 10&57213&1.21$\times 10^9$& $(1.0^{+0.2}_{-0.1})\times10^{-3}$& [6]\\[0.05cm]
I&July 11&57214&3.72$\times 10^5$& $22.3\pm2.0$& [8]\\[0.05cm]
R&July 11&57214&4.56$\times 10^5$& $21.1\pm1.1$& [8]\\[0.05cm]
V&July 11&57214&5.48$\times 10^5$& $15.0\pm0.4$& [8]\\[0.05cm]
Chandra&July 11&57214&1.21$\times 10^9$& $(1.2^{+0.02}_{-0.01})\times10^{-3}$& [7]\\[0.05cm]
I&July 12&57214&3.72$\times 10^5$& $20.1\pm1.0$& [8]\\[0.05cm]
R&July 12&57214&4.56$\times 10^5$& $17.6\pm1.6$& [8]\\[0.05cm]
V&July 12&57214&5.48$\times 10^5$& $13.7\pm0.8$& [8]\\[0.05cm]
\hline
\end{tabular}\\
\begin{flushleft}
{$^a$ This column displays the filters/instruments used for each observation; U (UVOT U band), B (UVOT B band), V (UVOT/Optical V band), UV1 (UVOT UVW1 band), UV2 (UVOT UVW2 band), R (optical R band), I (optical I band), J (infrared J band), H (infrared H band), K (infrared K band), {\em Swift} X-ray Telescope (XRT; 0.5-10 keV), {\em Chandra} (0.5-10 keV). Data shown in this table have been de-reddened (when required) using the prescription in \cite{cardelli89}, with an $E(B-V)=1.3\pm0.2$ \citep{cas93}.}\\
{$^b$ {\bf References:} [1] \citealt{ark15}; [2] \citealt{oat15}; [3] \citealt{shaw15}; [4] \citealt{carrascol15}; [5] \citealt{kim16};
[6] \citealt{sivakoffgr15d};
[7] \citealt{plot16}; [8] AAVSO; Kafka, S., 2018, Observations from the AAVSO International Database, \url{https://www.aavso.org}.}\\

\end{flushleft}
\label{table:other_flux}
\end{table*}
\renewcommand\tabcolsep{6pt}

\bsp	
\label{lastpage}
\end{document}